\documentclass[10pt,final]{IEEEtran}
\usepackage{amssymb,amsfonts}    
\usepackage{amsmath,amsthm}
\usepackage[pdftex]{graphicx}
\usepackage[latin1]{inputenc}

\usepackage{algorithmic}
\usepackage{cite}

\newtheorem{theorem}{Theorem}[section]
\newtheorem{proposition}[theorem]{Proposition}
\newtheorem{lemma}[theorem]{Lemma} 
\newtheorem{corollary}[theorem]{Corollary} 

\theoremstyle{definition} 
\newtheorem{definition}[theorem]{Definition} 
\newtheorem{remark}[theorem]{Remark}

\newtheorem{example}[theorem]{Example}

\newenvironment{sbmatrix}[1]{\left[\begin{array}{#1}}{\end{array}\right]}
\newcommand{\eqr}[1]{~\mbox{$(${\rm \ref{#1}}$)$}}
\newcommand{\dfree}{\mbox{$d_{\mbox{\rm\tiny free}}$}}
\newcommand{\F}{\mathbb{F}}
\def\wt{\operatorname{wt}}
\newcommand{\C}{\mathcal{C}}

\renewcommand{\vec}[1]{\ensuremath{\text{\mathversion{bold}$#1$}}}
\newcounter{abc}

\newcommand{\vect}[1]{\mathbf{#1}}
\newcommand{\length}{\ensuremath{\mathrm{length}}}
\newcommand{\rev}{\ensuremath{\mathrm{rev}}}

\newenvironment{alphalist}{\begin{list}{(\alph{abc})\hfill}{\usecounter{abc}
     \topsep.5ex \labelwidth.6cm \leftmargin.7cm \labelsep.1cm
     \rightmargin0cm \parsep0ex \itemsep.6ex
     \partopsep1.6ex}}{\end{list}}

\begin{document}

\title{Decoding of Convolutional Codes\\
  over the Erasure Channel}

\author{Virtudes Tom\'{a}s,
     Joachim Rosenthal,~\IEEEmembership{Senior Member,~IEEE},
    and Roxana Smarandache,~\IEEEmembership{Member,~IEEE}.

\thanks{V. Tom\'{a}s is with the Department of Computational
      Science and Artificial Intelligence University of Alicante
      Alicante, Spain. (e-mail: vtomas@dccia.ua.es). Her research has
      been supported by Spanish grant MTM2008-06674-C02-01 and a grant
      of the Vicerectorat d'Investigaci\'{o}, Desenvolupament i
      Innovaci\'{o} of the Universitat d'Alacant for PhD students
      during a stay at Z\"urich Universit\"at on charge to the same
      program. }

\thanks{J. Rosenthal is with the  Mathematics Institute, University of Z\"urich,
  Winterthurerstr 190,
    CH-8057 Z\"urich, Switzerland. (www.math.uzh.ch/aa). His research was supported 
  in part by the Swiss National Science
  Foundation under Project no. 126948.}

\thanks{R. Smarandache is with the Department of Mathematics and Statistics, 
    San Diego State University,
    San Diego, CA 92182-7720, USA.  (e-mail: rsmarand@sciences.sdsu.edu).
   Her research is supported in part by  by NSF Grants DMS-0708033 and
  TF-0830608.}

\thanks{Part of these results where presented
  in the 2009 IEEE International Symposium on Information Theory in
  Seoul, Korea~\cite{to09p} and the 2010 International Symposium on
  Mathematical Theory of Networks and Systems~\cite{to10p}.}
}

\maketitle

\begin{abstract}
  In this paper the decoding capabilities of convolutional codes over
  the erasure channel are studied. Of special interest will be maximum
  distance profile (MDP) convolutional codes. These are codes which
  have a maximum possible column distance increase.

  It is shown how this strong minimum distance condition of MDP
  convolutional codes help us to solve error situations that maximum
  distance separable (MDS) block codes fail to solve. Towards this
  goal, two subclasses of MDP codes are defined: reverse-MDP
  convolutional codes and complete-MDP convolutional codes.
  Reverse-MDP codes have the capability to recover a maximum number of
  erasures using an algorithm which runs backward in time.
  Complete-MDP convolutional codes are both MDP and reverse-MDP codes.
  They are capable to recover the state of the decoder under the
  mildest condition. It is shown that complete-MDP convolutional codes
  perform in many cases better than comparable MDS block codes of the
  same rate over the erasure channel.
\end{abstract}


\section{Introduction}

When transmitting over an erasure channel like the Internet, one of
the problems encountered is the delay experienced on the received
information due to the possible re-transmission of lost packets.  One
way to eliminate these delays is by using forward error
correction. Until now mainly block codes have been used for such a
task, see, e.g.,~\cite{fa08p,la04a} and the references therein. The use of
convolutional codes over the erasure channel has been studied much
less. We are aware of the work of Epstein~\cite{ep58} and of the more
recent work by Arai et al.~\cite{ar01}. In this paper, we define a
class of convolutional codes with strong distance properties, which we
call {\em complete maximum distance profile} (complete-MDP)
convolutional codes, and we demonstrate how they provide an attractive
alternative.

The advantage that convolutional codes have over block codes, which
will be exploited in our algorithms,  is the flexibility obtained
through the ``sliding window'' characteristic % cha\-rac\-te\-ris\-tic
of convolutional codes. 
% We work with the received symbols as a
%polynomial, so we look at it as if it was a complete sequence and we
%can slide up and down along it.  
The received information can be grouped 
%re\-cei\-ved in\-for\-ma\-tion can be grou\-ped in blocks or windows
in appropriate ways, depending on the erasure bursts, and then be
decoded by decoding the ``easy'' blocks first. This flexibility in
grouping information brings certain freedom in the handling of
sequences; we can split the blocks in smaller windows, we can overlap
windows and we can proceed to decode in a less strict order.  The
blocks are not fixed as in the block code case, i.e., they do not have
a fixed grouping of a fixed length.  We can slide along the
transmitted sequence and decide the place where we want to start our
decoding. In other words, we can adapt the process to the pattern of
erasures we receive. With this ``sliding window'' property of
convolutional codes, together with the extra algebraic properties of
maximum distance profile (MDP) convolutional codes, we are able to
correct in a given block more erasures than a block code of that same
length could do.

An $[N,K]$ block code used for transmission over an erasure
channel can correct up to $N-K$ erasures in a given block,  with the
optimal error capability of $N-K$ being  achieved by an $[N,K]$
maximum distance separable (MDS) code.

As an alternative, consider now a class of $(n,k,\delta)$
convolutional codes, i.e., a class of rate $k/n$ convolutional codes
having degree~$\delta$. We will demonstrate that for this class, the
maximum number of errors which can be corrected in some sliding window
of appropriate size is achieved by the subclass of MDP convolutional
codes.  Moreover, we give examples of situations where the MDP code
can recover patterns of erasures that cannot be decoded by an MDS
block code of the same rate. In addition, we can increase further the
recovering capability of MDP codes by imposing certain extra algebraic
conditions and thus defining a subclass of MDP convolutional codes,
called reverse maximum distance profile (reverse-MDP) convolutional
codes. These codes allow an inversion of the direction of the decoding
from right-to-left. Due to this fact one can recover through a
backward process more erasures than with an MDP code.  Following the
definition and explanation of their advantages, we will prove the
existence of the reverse-MDP codes and give a particular construction
as well as a procedure to compute the so called {\em
  reverse-superregular matrices} necessary to build them.

As a final step we add stronger and more restrictive conditions to our
codes in order to achieve an optimum performance of the recovering
process. We obtain what we call complete-MDP convolutional
codes. These codes help reducing the waiting time necessary when a
large burst of erasures occurs and no correction is possible for a
while.  Simulations results show that these codes can decode extremely
efficiently when compared to MDS block codes.  Thus, they provide a
very attractive choice when transmitting over an erasure channel.

MDP convolutional codes were first introduced in~\cite{hu05}. The
usefulness of the MDP property when transmitting over the erasure
channel was first recognized by the authors in two conference
papers~\cite{to09p,to10p}. The concept of complete MDP convolutional
codes was first introduced by the first author in her
dissertation~\cite{to10t}. The results we present here are an
extension of~\cite{to10t}.

Because of the increasing importance of packet switched networks the
need to develop coding techniques for the erasure channel has gained a
lot of importance. On the side of block codes and convolutional codes
of a fixed rate there have been important studies done using so called
``rateless erasure codes''. These codes were first introduced by
Luby~\cite{lu02} and an important refinement was done by Shokrollahi
who introduced so called Raptor codes~\cite{sh06a}. In this paper we
will not make a performance comparison of MDP codes with rateless
erasure codes.

The paper is organized as follows. Section~\ref{Preliminaries}
provides the necessary background for the development of the
paper: Subsection \ref{Erasure} explains the assumptions on the
channel model, and Subsection~\ref{MDP} provides all the necessary
concepts about convolutional codes, MDP convolutional codes and their
characterizations. Section~\ref{Decoding} illustrates our proposed
decoding algorithm over the erasure channel. It also presents examples
and special concerns to be addressed when comparing them with MDS
block codes. In Section~\ref{back}, we introduce the idea of backwards
decoding process and we define and prove the existence of reverse-MDP
convolutional codes as codes able to do this. In
Section~\ref{ConstRev}, we give a method to construct these codes and
in Subsection~\ref{ConstMat}, we explain how to construct a special
kind of matrices necessary in order to build reverse-MDP convolutional
codes.  Section~\ref{Stronger} introduces the concept of complete-MDP
convolutional codes and shows how these codes can help to reduce the
waiting time in the recovering process. In
Subsection~\ref{Simulations} we provide simulation result assuming 
a Gilbert-Elliot channel model. It is shown that for equal rate 
and chosen degrees comparable to chosen block length the performance 
of complete-MDP codes are a better  option than MDS block codes.

In Section~\ref{Comparison} we provide theoretical results which
compare MDS block codes with MDP convolutional codes. The main result
shows that both a rate $k/n$ MDS block code as well as a $k/n$ MDP
convolutional code can decode erasures at a rate of $(n-k)/n$ in
average.  For the MDS block code error free communication is possible
if at most $n-k$ erasures happen in every block. For the MDP
convolutional code error free communication is possible if the number
of erasures per sliding window, whose size depends on the degree, is
not larger than a certain amount.

%%%%%%%%%%%%%%%%%%%%%%%%%%%%%%%%%%%%%%%%%%%%%%%%%%%%%%%%%%%%%%%%%%
\section{Preliminaries} \label{Preliminaries}
  This section contains  the necessary mathematical
  background and the channel assumptions needed for the
  development of our results. Note that throughout the paper vectors
  of length $n$ over a field $\F$ will be viewed as $n\times 1$
  matrices, i.e., as  column vectors.
\subsection{Erasure channel} \label{Erasure}
	
An erasure channel is a communication channel where the symbols sent
either arrive correctly or they are erased; the receiver knows that a
symbol has not been received or was received incorrectly. An important
example of an erasure channel is the Internet, where packet sizes are
upper bounded by 12,000 bits - the maximum that the Ethernet protocol
allows (that everyone uses at the user end).
In many cases, this
maximum is actually used \cite{fr03}. 
Due to the nature of the TCP part of the
TCP/IP protocol stack, most sources need an acknowledgment confirming
that the packet has arrived at the destination; these packets are only
320 bits long. So if everyone were to use TCP/IP, the packet size
distribution would be as follows: 35\% --320 bits, 35\% -- 12,000 bits
and 30\% -- uniform distribution in between the two.  Real-time
traffic used, e.g., in video calling, does not need an acknowledgment
since that would take too much time; overall, the following is a good
assumption of the packet size distribution: 30\% -- 320 bits, 50\% --
12,000 bits, 20\% --uniform distribution in between, see \cite{si07r}
and \cite[Table II]{le08}.

We can model each packet as an element or sequence of elements from a
large alphabet. Packets sent over the Internet are protected by a
cyclic redundancy check (CRC) code. If the CRC check fails, the
receiver knows that a packet is in error or has not arrived
\cite{pa99a}; it then declares an erasure.  Undetected errors are rare
and are ignored. For illustration purpose we employ as alphabet the
finite field $\F:=\F_{2^{1,000}}$. If a packet has less than 1,000
bits, then one uses simply the corresponding element of~$\F$. If the
packet is larger, one uses several alphabet symbols to describe the
packet. With or without interleaving, such an encoding scheme results
in the property that errors tend to occur in bursts, and this is a
phenomenon observed about many channels modeled via the erasure
channel.  This point is important to keep in mind when designing codes
which are capable of correcting many errors over the erasure channel.
	
\subsection{Convolutional codes} \label{MDP}

Let $\mathbb{F}$ be a finite field.  We view a convolutional code
$\mathcal{C}$ of rate $k/n$ as a submodule of $\mathbb{F}[z]^{n}$
(see \cite{gl08,ro96a1,ro01}) that can be described as
\[
\mathcal{C}=\left\{ \vec{v}(z) \in \mathbb{F}[z]^{n} \ | \
  \vec{v}(z)=G(z)\vec{u}(z) \ \text{with} \ \vec{u}(z) \in
  \F[z]^{k} \right\},
\]
where $G(z)$ is an $n\times k$ full-rank polynomial matrix called a
\textbf{generator matrix} for $\mathcal{C}$, $\vec{u}(z)$ is an
\textbf{information vector}, and $\vec{v}(z)$ is the resulting
\textbf{code vector} or the \textbf{codeword}.

The maximum degree of all polynomials in the $j$-th column of $G(z)$
is called the $j$-th \textbf{column degree} of $G(z)$, and we denote
it by $\delta_j$.

We define the \textbf{degree} $\delta$ of a convolutional code
$\mathcal{C}$ as the maximum of the degrees of the determinants of the
$k\times k$ sub-matrices of one and  hence any generator matrix of
$\mathcal{C}$.  We say that $\mathcal{C}$ is an $(n,k,\delta)$
convolutional code~\cite{mc98}.

Assume the $j$-th column of $G(z)$ has degree $\delta_j$.  The high
order coefficients matrix of $G(z)$, $G_{\infty}$, is the matrix whose
$j$-th column is formed by the coefficients of $z^{\delta_j}$ in the
$j$-th column of $G(z)$. If $G_{\infty}$ has full rank, then $G(z)$ is
called a \textbf{minimal generator matrix} and the degree $\delta$ of
the code agrees in this situation with the overall constraint length
(see~\cite[Section 2.5]{jo99}) of the encoder $G(z)$. Note that in
this case $\delta=\sum_{i=1}^{k}\delta_i$. Finally we define the
\textbf{memory} of an encoder $G(z)$ as the maximum of the column
degrees $\{\delta_1,\ldots,\delta_k\}$. This is the parameter of an
encoder.  When we choose however a minimal generator matrix then the
memory becomes the property of the convolutional code.

We say that a code $\mathcal{C}$ is \textbf{observable} (see, e.g.,
\cite{ro99a,ro01}) if the generator matrix $G(z)$ has a polynomial
left inverse. This avoids the type of catastrophic situations in which
a sequence $\vec{u}(z)$ with an infinite number of nonzero
coefficients can be encoded into a sequence $\vec{v}(z)$ with a finite
number of nonzero coefficients; this case would decode finitely many
errors on the received code sequence into infinitely many errors when
recovering the original information sequence. Therefore, only
observable codes are regularly considered; therefore, these will be
the codes on which we will focus our attention.

If $\C$ is an observable code, then it can be equivalently
described using an $(n-k)\times n$ full rank polynomial
\textbf{parity-check matrix} $H(z)$, such that
\[
\mathcal{C}=\left\{ \vec{v}(z) \in \mathbb{F}[z]^{n} \ \ | \ \
  H(z)\vec{v}(z)=\vec{0} \in \mathbb{F}[z]^{n-k} \right\}.
\]
If we write
$\vec{v}(z)=\vect{v}_{0}+\vect{v}_{1}z+\ldots+\vect{v}_{l}z^{l}$,
(with $l\geq 0$), and we represent $H(z)$ as a matrix polynomial, 
\[
H(z)=H_{0}+H_{1}z+\cdots+H_{\nu}z^{\nu},
\]
where $H_{i}=O$, for $i > \nu$, we can expand the kernel
representation in the following way
\begin{eqnarray}\label{eq0}
  \begin{sbmatrix}{ccccc}
    H_{0}   &   \\
    \vdots  & \ddots &  \\
    H_{\nu} & \cdots & H_{0}    \\
    & \ddots &         & \ddots  \\
    &        & H_{\nu} & \cdots & H_{0} \\
    &        &         & \ddots & \vdots \\
    &        &         &        & H_{\nu} \\
  \end{sbmatrix}
  \begin{sbmatrix}{c}
    \vect{v}_{0} \\
    \vect{v}_{1} \\
    \vdots \\
    \vect{v}_{l}
  \end{sbmatrix}=\vec{0}.
\end{eqnarray}

An important distance measure for convolutional codes is the
\textbf{free distance} $\dfree$ defined as
\[
\dfree(\mathcal{C}):=\min \left\{ \wt(\vec{v}(z)) \ \ | \ \
  \vec{v}(z) \in \mathcal{C} \quad \text{and} \quad
  \vec{v}(z)\neq \vec{0} \right\}.
\]
The following lemma shows the importance of the free distance as
a performance measure of a code used over the erasure channel.
\begin{lemma}
  Let $\C$ be a convolutional code with free distance
  $d\stackrel{\text{def}}{=}\dfree$. If during the transmission at
  most $d-1$ erasures occur, then these erasures can be uniquely
  decoded. Moreover, there exist patterns of $d$ erasures which cannot
  be uniquely decoded. \end{lemma}
\begin{IEEEproof} Let
  $\vec{v}(z)=\vect{v}_{0}+\vect{v}_{1}z+\ldots+\vect{v}_{l}z^{l}$ be
  a received vector with $d-1$ erased symbols erased in positions
  $i_1, \ldots, i_{d-1}$. The homogeneous system~\eqref{eq0} of
  $(\nu+l+1)(n-k)$ equations with $(l+1)n$ unknowns can be changed
  into an equivalent non-homogeneous system
  \begin{eqnarray}\label{syst}
    \hat H 
    \begin{sbmatrix}{c}
      v_{i_1}\\
      v_{i_2}\\
      \vdots \\
      v_{i_{d-1}}
    \end{sbmatrix}
    =\vect{b} 
  \end{eqnarray}
  of $(\nu+l+1)(n-k)$ equations with $d-1$ unknowns $v_{i_1}, \ldots,
  v_{i_{d-1}}$ where $\hat H$ is a $(d-1)(n-k)\times(d-1)n$ sub-matrix
  of
  \[
  H=\begin{sbmatrix}{ccccc}
    H_{0}   &   \\
    \vdots  & \ddots &  \\
    H_{\nu} & \cdots & H_{0}    \\
    & \ddots &         & \ddots  \\
    &        & H_{\nu} & \cdots & H_{0} \\
    &        &         & \ddots & \vdots \\
    &        &         &        & H_{\nu} \\
  \end{sbmatrix}.
  \]
  This non-homogeneous system (\ref{syst}) has a solution, because of
  the assumption that the channel allows only erasures. In addition,
  the columns of the system matrix are linearly independent, because
  $d=\dfree(\mathcal{C})$, so the matrix $\hat H$ has full column
  rank. It follows from these two facts that the solution must be
  unique.

  If on the other hand more than $d$ erasures happen, then the
  associated linear system of equations does not have a unique
  solution anymore.
\end{IEEEproof}

Rosenthal and Smarandache~\cite{ro99a1} showed that the free distance
of an $(n,k,\delta)$ convolutional code must be upper bounded by
\begin{eqnarray}\label{eq1}
  \dfree(\mathcal{C})\leq (n-k)\left(\left\lfloor 
      \frac{\delta}{k}\right\rfloor + 1\right)+\delta +1.
\end{eqnarray}
This bound is known as the \textbf{generalized Singleton bound}
\cite{ro99a1} since it generalizes in a natural way the Singleton
bound for block codes (the case $\delta=0$). Moreover, an
$(n,k,\delta)$ convolutional code is a \textbf{maximum distance
  separable} (MDS) code~\cite{ro99a1} if its free distance achieves
the generalized Singleton bound.

Another important distance measure is the \textbf{$j$th column distance}
\cite{jo99}, $d_{j}^{c}(\mathcal{C})$, given by the expression
$$ d_{j}^{c}(\mathcal{C})=\min\left\{ \wt(\vect{v}_{[0,j]}(z))\ \ |
\ \ \vec{v}(z)\in \mathcal{C} \ \ \text{and} \ \ \vect{v}_{0}\neq
\vec{0}\right\},
$$ where
$\vect{v}_{[0,j]}(z)=\vect{v}_{0}+\vect{v}_{1}z+\ldots+\vect{v}_{j}z^{j}$
represents the $j$-th truncation of the codeword $\vec{v}(z)\in
\mathcal{C}$.  It is related to the free distance
$\dfree(\mathcal{C})$ in the following way
\begin{eqnarray}\label{eq2}
  \dfree(\mathcal{C})=\lim_{j\rightarrow \infty} d_{j}^{c}(\mathcal{C}).
\end{eqnarray}
The $j$-th column distance is upper bounded \cite{gl06,hu05}
\begin{eqnarray}\label{eq3}
  d_{j}^{c}(\mathcal{C})\leq (n-k)(j+1)+1,
\end{eqnarray}
and the maximality of any of the column distances implies the
maximality of all the previous ones, i.e., if
$d_{j}^{c}(\mathcal{C})=(n-k)(j+1)+1$ for some $j$, then
$d_{i}^{c}(\mathcal{C})=(n-k)(i+1)+1$ for all $i\leq j$, see
\cite{gl06,hu05}.  The $(m+1)$-tuple
$(d_{0}^{c}(\mathcal{C}),d_{1}^{c}(\mathcal{C}),\ldots,d_{m}^{c}(\mathcal{C}))$
is called the \textbf{column distance profile} of the code
\cite{jo99}.

Since no column distance can achieve a value greater than the
generalized Singleton bound, there must exist an integer $L$ for which
the bound (\ref{eq3}) could be attained for all $j\leq L$ and it is a
strict upper bound for $j>L$; this value is
\begin{eqnarray}\label{eq4}
  L=\left\lfloor \frac{\delta}{k} \right\rfloor + 
  \left\lfloor \frac{\delta}{n-k} \right\rfloor.
\end{eqnarray}
An $(n,k,\delta)$ convolutional code $\mathcal{C}$ with
$d_{L}^{c}(\mathcal{C})=(n-k)(L+1)+1$ is called a \textbf{maximum
  distance profile} (MDP) code~\cite{gl06,hu05}.  In this case, every
$d_{j}^{c}(\mathcal{C})$ for $j\leq L$ is maximal, so we can say that
the column distances of MDP codes increase as rapidly as possible for
as long as possible.

The following two theorems characterize algebraically all
convolutional codes of a given $j$th column distance $d$, and hence
also MDP convolutional codes. Assume that the parity-check matrix is
given as $H(z)=\sum^{\nu}_{i=0} H_{i}z^{i}$. For each $j>\nu$, let
$H_j=O$ and define:
\begin{equation} \label{eq5} \mathcal{H}_{j}=
  \begin{sbmatrix}{cccc}
    H_{0}  &         &        & \\
    H_{1}  & H_{0}   &        & \\
    \vdots & \vdots  & \ddots & \\
    H_{j} & H_{j-1} & \cdots & H_{0}
  \end{sbmatrix}
  \in \mathbb{F}^{(j+1)(n-k)\times(j+1)n}, 
\end{equation}
for all $j\geq 0$.
\begin{theorem}(\cite[Proposition 2.1]{gl06})\label{Theorem-d}
  Let $d\in\mathbb{N}$. The following pro\-per\-ties are
  equivalent.
  \begin{alphalist}
  \item $d^c_j=d$;
  \item none of the first~$n$ columns of~$\mathcal{H}_{j}$ is
    contained in the span of any other $d-2$ columns and one of the
    first~$n$ columns of~$\mathcal{H}_{j}$ is in the span of some
    other $d-1$ columns of that matrix.
  \end{alphalist}
\end{theorem}
Let $G(z)=\sum_{i=0}^{m}G_iz^i$, $G_j=O,$  for all $j>\nu$,  and
\begin{equation}\label{eq9}
  \mathcal{G}_j=\begin{sbmatrix}{cccc}
    G_0 & G_1 & \cdots & G_j \\
    & G_0 & \cdots & G_{j-1} \\
    &     & \ddots & \vdots \\
    &     &        & G_0
  \end{sbmatrix}, \text{~for all ~} j\geq 0.
\end{equation}
Then, the  MDP convolutional codes are
characterized as follows:
\begin{theorem}(\cite[Theorem 2.4]{gl06})\label{mdp} Let
  $\mathcal{G}_j$ and $\mathcal{H}_j$ be like in (\ref{eq9}) and
  (\ref{eq5}). Then the following are equivalent:
  \begin{alphalist}
  \item \label{Thgen1} $d_j^c=(n-k)(j+1)+1$;
  \item \label{Thgen2} every $(j+1)k\times (j+1)k$ full-size
    minor of $\mathcal{G}_j$ formed from the columns with indices
    $1\leq t_1 < \cdots < t_{(j+1)k}$, where $t_{sk+1}>sn$, for
    $s=1,2,\ldots,j$, is nonzero;
  \item \label{Thgen3} every $(j+1)(n-k)\times (j+1)(n-k)$
    full-size minor of $\mathcal{H}_j$ formed from the columns
    with indices $1\leq r_1 < \cdots < r_{(j+1)(n-k)}$, where
    $r_{s(n-k)}\leq sn$, for $s=1,2,\ldots,j$, is nonzero.
  \end{alphalist}
In particular, when $j=L$, $\mathcal{C}$ is an MDP convolutional code.
\end{theorem}
A code satisfying the conditions of 
Theorem~\ref{mdp} is said to have the \textit{MDP property}.

Note that MDP convolutional codes are similar to MDS block codes
within windows of size $(L+1)n$.  Indeed, the nonsingular full-size
minors property given in the previous theorem ensures that if we
truncate a codeword with its first nonzero component  at any $j$
component, with $j\leq L$, it will have weight higher or equal than
the bound given in Theorem \ref{mdp} \textit{(a)}, which is the
Singleton bound for that block code.

\section{Decoding over an erasure channel} \label{Decoding}

Let us suppose that we use a convolutional code $\mathcal{C}$ to
transmit over an erasure channel.  Then we can state the
following result.
\begin{theorem} 
  \label{main2} Let $\mathcal{C}$ be an $(n,k,\delta)$ convolutional
  code with $d_{j_{0}}^{c}$ the $j_{0}$-th column distance.  If in any
  sliding window of length $(j_{0}+1)n$ at most $d_{j_{0}}^{c}-1$
  erasures occur, then we can completely recover the transmitted
  sequence.
\end{theorem}

\begin{IEEEproof} Assume that we have been able to correctly decode up
  to an instant $t-1$. Then we have the following homogeneous
  system: \small
  \begin{align}\label{eqdec}
    \begin{sbmatrix}{cccccccccc}
      \!\! H_{\nu} \!\!\!\!\! & H_{\nu-1} \!\!\!\!\! & \cdots 
    \!\!\!\!\!  &  H_{\nu -j_{0}}  \!\!\!\!\!  & \cdots  
    \!\!\!\!\! &  H_{0}   \!\!\!\!\!   \\
      \!\!\!\!\!  & H_{\nu}   \!\!\!\!\! & \cdots
    \!\!\!\!\!  &  H_{\nu -j_{0}+1} \!\!\!\!\! & \cdots    
    \!\!\!\!\! &  H_{1}    \!\!\!\!\!  & H_{0}       \!\!\!\!\! \\
      \!\!\!\!\!  &           \!\!\!\!\! & \ddots 
    \!\!\!\!\!  &                   \!\!\!\!\! &       
      \!\!\!\!\! &           \!\!\!\!\!  &              \!\!\!\!\! & 
     \ddots \!\!\!\!\!  \\
      \!\!\!\!\! & \!\!\!\!\! & \!\!\!\!\!  & H_{\nu} \!\!\!\!\!
      & \cdots \!\!\!\!\! & H_{j_{0}} \!\!\!\!\! & H_{j_{0}-1}
      \!\!\!\!\! & \cdots \!\!\!\!\!  & H_{0} \!\!
    \end{sbmatrix}
    \begin{sbmatrix}{c}
      \!\!\! \vect{v}_{t-\nu} \!\!\! \\
      \!\!\!  \vdots \!\!\! \\
      \!\!\! \vect{v}_{t-1} \!\!\! \\
      \!\!\! \star \!\!\! \\
      \!\!\! \star \!\!\! \\
      \!\!\! \vdots \!\!\! \\
      \!\!\! \star \!\!\!
    \end{sbmatrix}=\vec{0},
  \end{align}
  \normalsize where $\star$ takes the place of a vector that had
  some of the components erased. Let the positions of the erased
  field elements be $i_1, \ldots, i_{e},$ $e\leq
  d_{j_{0}}^{c}-1$, where $i_1, \ldots, i_{s},$ $s\leq n$, are
  the erasures occurring in the first $n$-vector erased.  We can
  take the columns of the matrix in equation (\ref{eqdec}) that
  correspond to the coefficients of the erased elements to be the
  coefficients of a new system. The rest of the columns in
  (\ref{eqdec}) will help us to compute the independent terms. In
  this way we get a non-homogeneous system with $(j_{0}+1)(n-k)$
  equations and $e\leq d_{j_{0}}^{c}-1$, variables.
 
  We claim that there is an extension $\{ \vect{\tilde v}_{t},
  \ldots, \vect{\tilde v}_{t+j_{0}}\}$ such that the vector 
  \[  (\vect{v}_{t-\nu}, \ldots, \vect{v}_{t-1}, \vect{\tilde v}_{t},
  \ldots, \vect{\tilde v}_{t+j_{0}})
  \]
   is a codeword and such that
  $\vect{\tilde v}_{t}$ is unique.

  Indeed, we know that a solution of the system exists since we
  assumed that only erasures occur.  To prove the uniqueness of
  $\vect{\tilde v}_{t}$, or equivalently, of the erased elements
  $\tilde v_{i_1}, \ldots, \tilde v_{i_{s}},$ let us suppose
  there exist two such good extensions $\{ \vect{\tilde v}_{t},
  \ldots, \vect{\tilde v}_{ t+j_{0}}\}$ and $\{
  \vect{\tilde{\tilde v}}_{t}, \ldots, \vect{\tilde{\tilde v}}_{
    t+j_{0}}\}$.  Let $\vect{h}_{i_1}, \ldots, \vect{h}_{i_e}$ ,
  be the column vectors of the sliding parity-check matrix
  in~\eqref{eqdec} which correspond to the erasure elements. We
  have:
$$ \tilde
v_{i_1} \vect{h}_{i_1} + \cdots +{\tilde
  v}_{i_{s}}\vect{h}_{i_s}+\cdots + \tilde
v_{i_{e}}\vect{h}_{i_e}=\tilde{\vect{b}}$$ and
$$
\tilde{\tilde v}_{i_1} \vect{h}_{i_1} + \cdots +\tilde{\tilde
  v}_{i_{s}}\vect{h}_{i_s} + \cdots +\tilde{\tilde
  v}_{i_{e}}\vect{h}_{i_e}= \tilde{\tilde{\vect{b}}},$$ where the
vectors $\tilde{\vect{b}}$ and $\tilde{\tilde{\vect{b}}}$
correspond to the known part of the system.  Subtracting these
equations and observing that
$\tilde{\vect{b}}=\tilde{\tilde{\vect{b}}}$, we obtain: {\small
  $$
  (\tilde v_{i_1}-\tilde{\tilde v}_{i_1}) \vect{h}_{i_1} + \cdots
  +(\tilde v_{i_{s}}- \tilde{\tilde
    v}_{i_{s}})\vect{h}_{i_s}+\cdots + (\tilde v_{i_{e}}-
  \tilde{\tilde v}_{i_{e}})\vect{h}_{i_e}=\vec{0}.
$$}
Using Theorem~\ref{Theorem-d} part (b) we obtain that,
necessarily,
\begin{align*} \tilde v_{i_1}-&\tilde{\tilde
    v}_{i_1}=0,%\\& ~~~~~~\vdots
  ~ \ldots, ~\tilde v_{i_{s}}-%&
  \tilde{\tilde v}_{i_{s}}=0, \end{align*}
 which proves the uniqueness of the solution.

In order to find the value of this unique vector, we solve the
full column rank system, find a solution and retain the part
which is unique. Then we slide $n$ bits to the next $n(j_{0}+1)$
window and proceed as above.
\end{IEEEproof}

The best scenario of Theorem~\ref{main2} happens when the
convolutional code is MDP.  In this case, full error correction `from
left to right' is possible as soon as the fraction of erasures is not
more than $\frac{n-k}{n}$ in any sliding window of length $(L+1)n$.

\begin{corollary} \label{main} Let $\mathcal{C}$ be an
  $(n,k,\delta)$ MDP convolutional code.  If in any sliding
  window of length $(L+1)n$ at most $(L+1)(n-k)$ erasures occur
  in a transmitted sequence, then we can completely recover the
  sequence in polynomial time in $\delta$ by iteratively decoding 
the symbols `from left to right'.
\end{corollary}
\begin{IEEEproof}
Under the given assumptions it is possible to compute one
erasure after the other in a unique manner by processing `from
left to right'.
\end{IEEEproof}

%\subsection{Examples and Remarks}\label{example}

\begin{remark}                   \label{rem-algo}
  The process of computing the erasures described in the proof of
  Theorem~\ref{main2} leads to a natural algorithm.  The computation
  of each erased symbol requires only simple linear algebra. In the
  optimum case of an MDP convolutional code, for every set of
  $(L+1)(n-k)$ erasures, a matrix of size at most $(L+1)(n-k)$ has to
  be inverted over the base field~$\F$. This is easily achieved even
  over fairly large fields. To be precise the number of elementary
  field operations is $\mathcal{O}\left(L^3(n-k)^3\right)$ and the
  most costly field operation in $\F_q$, namely division, requires
  $\mathcal{O}\left(\log^3 q\right)$ bit operations.  \hfill$\square$
\end{remark}

\begin{remark}
  Theorem~\ref{main} is optimal in the following sense.  One can show
  that for any $(n,k,\delta)$ code there exist patterns of
  $(L+2)(n-k)$ erasures in a sliding window of length $(L+2)n$
  which cannot be uniquely decoded.  

  In Corollary~\ref{rec-rate} we will show that the maximal recovering
  rate of any rate $k/n$ convolutional code over the erasure channel
  is at most $R=\frac{n-k}{n}$.  \hfill$\square$
\end{remark}

\begin{remark}\label{remark1}
  Although in Theorem~\ref{main2} we fix the value $j=j_{0}$, other
  window sizes can be taken during the decoding process in order to
  optimize it. For any value of $j$, at most $d_{j}^{c}-1$ erasures
  can be recovered in a window of size $(j+1)n$. In the MDP case, the
  parameter $L$ gives an upper bound on the length of the window we
  can take to correct. For every $j\leq L$, in a window of size
  $(j+1)n$ we can recover at most $(j+1)(n-k)$ erasures. This means
  that we can conveniently choose the size of the window we need at
  each step depending on the distribution of the erasures in the
  sequence. This is an advantage of these codes over block codes. If
  we receive a part of sequence with a few errors we do not need to
  wait until we receive the complete block, we can already proceed
  with decoding within small windows relative to $L$.
  
  This property allows us to recover the erasures in situations
  where the MDS block codes cannot do it. The following example
  illustrates this scenario. We compare an MDP convolutional code
  with an MDS block code of the same length as the maximum window
  size taken for the convolutional code.  \hfill$\square$
\end{remark}

\begin{example}\label{examp1}
  Consider a $(2,1,50)$ MDP convolutional code over an erasure
  channel.  In this case, the decoding can be completed if in any
  sliding window of length $202$ not more than $101$ erasures occur;
  therefore, $50\%$ of the erasured components can be correctly
  recovered.
  
  An  MDS block code which can achieve a comparable performance is a
  $[202,101]$ MDS block code.  In a block of $202$ symbols we can
  recover $101$ erased symbols, which is again $50\%$ error capability.

  Suppose now that we have been able to correctly decode up to an
  instant $t$. After time $t$ a new block of symbols starts whose start
  is indicated by $|$. Assume now we receive the following pattern of
  erasures
  \[
  \ldots \vect{v}\vect{v}|\overbrace{\star \star \ldots \star
    \star}^{(A)60} \overbrace{\vect{v}\vect{v}\ldots
    \vect{v}}^{(B)80} \overbrace{\star \star \ldots \star
    \star}^{(C)60}\vect{v}\vect{v}|\vect{v}\vect{v}\ldots,
  \]
  where each $\star$ stands for a component of the vector that
  has been erased and $\vect{v}$ means that the component has
  been correctly received.  In this situation, $120$ erasures
  happen in a block of $202$ symbols making the  MDS block code unable 
  to recover them.  In the block code situation one has
  to skip the whole window and lose a whole block,  and move to the 
  next block.

  The MDP convolutional code proves to be a better choice in this
  situation.  If we frame a $120$ symbols length window, then in this
  window we can correct up to $60$ erasures. Let us frame a window
  containing the first $60$ erasures from $A$ and $60$ more correct
  symbols from $B$. Note that following expression (\ref{eqdec}), in
  order to solve the corresponding system and to help us calculate the
  independent terms, we need to take the $100$ correct symbols that we
  decoded before receiving the block $A$.
  % We can take $100$
%  previously decoded symbols, then frame a window containing the first
%  $60$ erasures from $A$ and $60$ more clean symbols from $B$. 
In  this way we can solve the system and recover the first block of $60$ erasures. 
  \[
  \overbrace{\vect{v}\vect{v}\ldots
    \vect{v}\vect{v}}^{100}|\overbrace{\star \star \ldots \star
    \star}^{(A)60}\overbrace{\vect{v}\vect{v}\ldots
    \vect{v}}^{(B)60}
  \]
  Then we slide through the received sequence until we frame the rest
  of the erasures in a $120$ symbols window. As before, we make use of
  the $100$ previously decoded symbols to compute the independent
  terms of the system.
  \[
  \overbrace{\vect{v}\vect{v}\ldots
    \vect{v}\vect{v}}^{(A+B)100}\overbrace{\star \star \ldots
    \star
    \star}^{(C)60}\overbrace{\vect{v}\vect{v}|\vect{v}\vect{v}
    \ldots}^{60}
  \]
  After recovering this block we have correctly decoded the sequence.
  \hfill$\square$
\end{example}

\begin{remark}               \label{remark3}
  There are situations in which other patterns of erasures than the
  ones covered by Theorem~\ref{main2} or Corollary~\ref{main} occur,
  and for which decoding within smaller window sizes than maximum
  allowed is not possible. This leads to an inability of correcting that
  block;  we say that we are {\em lost in the recovering process}.

  Looking at the following system of
  equations, %When using the parity-check matrix we know that
  %\small
  \begin{eqnarray*}
    \begin{sbmatrix}{cccccccccc}            
      H_{\nu} \!\!\!\!\! & H_{\nu-1} \!\!\!\!\! & \cdots \!\!\!\!\! & 
    H_{\nu -j}   \!\!\!\!\! & \cdots   \!\!\!\!\! &  H_{0} \!\!\!\!\! \\
      \!\!\!\!\! & H_{\nu}   \!\!\!\!\! & \cdots \!\!\!\!\! &  H_{\nu -j+1} 
     \!\!\!\!\! & \cdots   \!\!\!\!\! &  H_{1} \!\!\!\!\! & H_{0}  \!\!\!\!\! \\
      \!\!\!\!\! &           \!\!\!\!\! & \ddots \!\!\!\!\! &              
      \!\!\!\!\! &          \!\!\!\!\! &        \!\!\!\!\! &  
        \!\!\!\!\!  &  \ddots \!\!\!\!\! \\
      \!\!\!\!\! & \!\!\!\!\! & \!\!\!\!\! & H_{\nu} \!\!\!\!\! &
      \cdots \!\!\!\!\! & H_{j} \!\!\!\!\! & H_{j-1} \!\!\!\!\! &
      \cdots \!\!\!\!\! & H_{0}
    \end{sbmatrix}
    \begin{sbmatrix}{c}
      \!\!\! \vect{v}_{t-\nu} \!\!\! \\
      \!\!\! \vdots \!\!\! \\
      \!\!\! \vect{v}_{t} \!\!\! \\
      \!\!\! \vect{v}_{t+1} \!\!\! \\
      \!\!\! \vdots \!\!\! \\
      \!\!\! \vect{v}_{t+j} \!\!\!
    \end{sbmatrix}=\vec{0},
  \end{eqnarray*}
  \normalsize we see that, in order to continue our recovering
  process, we need to find a block of $\nu n$ correct symbols
  $\vect{v}_{t-\nu}$ to $\vect{v}_{t-1}$ preceding a block of $(j+1)
  n$ symbols $\vect{v}_{t}$ to $\vect{v}_{t+j}$ where not more than
  $d_{j}^{c}-1=(j+1)(n-k)$ erasures occur. In other words, we need to
  have some \textit{guard space}, an expression often used in the
  literature. (See e.g.~\cite[p. 288]{ga68} or~\cite[p. 430]{li83}).
  This allows a restart of the decoding algorithm leading to recovery
  of $\vect{v}_{t}$ to $\vect{v}_{t+j}$.

  In Section~\ref{Stronger} we will derive Theorem~\ref{complete-thm}
  which provides somehow the weakest conditions possible which will
  guarantee the computation of a guard space once the decoder is lost
  in the decoding process.

  We define the \textit{recovering rate per window} as
  $R_{\omega}=\frac{\# \text{erasures recovered}}{\# \text{symbols in
      a window}}$. Note that above condition of having  a ``guard
  space'' and restarting the recovering process is a sufficient
  condition for $R_{\omega}$ to be maintained. For any generic
  $(n,k,\delta)$ convolutional code,
  $R_{\omega}=\frac{d_{j}^{c}-1}{(j+1)n}$.  In the MDP case where the
  number of possible recovered erasures is maximized, we have
  $R_{\omega}=\frac{(j+1)(n-k)}{(j+1)n}$.  \hfill$\square$
\end{remark}

\section{The backward process and the reverse-MDP convolutional
  codes}\label{back}

In this section we define a subclass of MDP codes, called reverse-MDP
convolutional codes, which have the MDP property not only forward but
also backward, i.e., if we truncate sequences
$[\vect{v}_0,\ldots,\vect{v}_M],$ $M\geq L$, with $\vect{v}_0\neq \vec{0}$
and $\vect{v}_M\neq \vec{0}$ either at the beginning, to obtain
$[\vect{v}_0,\ldots,\vect{v}_L]$, or at the end, to obtain
$[\vect{v}_M,\ldots,\vect{v}_{M-L}]$, the minimum possible weight of
the segments obtained is as large as possible. We will see in the
following how this backward decoding ability of reverse-MDP codes makes
these codes better choices than regular MDP convolutional codes for
transmission over an erasure channel, since they can recover certain situations
in which the latter would fail.

% In Remark~\ref{remark3} we showed what are the necessary conditions
% needed in order to restart recovering once we get lost: once the
% number of erasures is too large for a given window, we need to find a
% window of correct or ``almost'' correctly transmitted symbols.
% Finding a whole block of correct symbols could however end up in a long
% waiting time.  This, together with the fact that a block or more of
% the sequence could be lost, constitutes the main weakness of this
% algorithm.  The following example illustrates this situation.

\begin{example}\label{ex4}
  As previously, assume we use a $(2,1,50)$ MDP convolutional code to
  transmit over an erasure channel. Suppose that we are able to
  recover the sequence up to an instant $t$, after which we receive a
  part of a sequence with the following pattern
  \[
  \ldots\vect{v}\vect{v} \overbrace{\star\ldots\star}^{(A)22} \
  \overbrace{\vect{v}\vect{v}\star\star \vect{v}\vect{v} \star\star
    \ldots \vect{v}\vect{v}\star\star}^{(B)180} \ |
  \overbrace{\vect{v}\vect{v}\ldots \vect{v}\vect{v}}^{(C)202} |
  \]
  \[
  | \ \overbrace{\star\star\ldots\star}^{(D)80} \
  \overbrace{\vect{v}\vect{v}\ldots \vect{v}}^{(E)62} \
  \overbrace{\star\star\ldots\star}^{(F)60} \ | \
  \overbrace{\vect{v}\vect{v} \ldots \vect{v}}^{(G)202}, 
  \]
  where, as before, $\star$ means that the symbol has been erased, and
  $\vect{v}$ denotes a symbol has been correctly received. This is a
  situation in which we cannot recover the sequence by simply decoding
  `from left to right' through the algorithm explained in
  remark~\ref{rem-algo}. The simple `from left to right' decoding
  algorithm for MDP convolutional codes needs to skip over these
  erasures, leading to the loss of this information.  A $[202,101]$
  MDS block code would not be a better choice either since in a block
  of $202$ symbols there would be more than $101$ erasures making that
  block undecodable.  \hfill$\square$
\end{example}

This example shows that even with enough guard space between bursts
of erasures, we cannot always decode if the bursts are too large
relative to a given window. Let us imagine the following scenario.  In
the places where a guard space appears we change our decoding direction
from left-to-right to right-to-left. Suppose that we could split the
sequence into windows starting from the end, such that erasures are
less accumulated in those windows, i.e., such that reading the
patterns right-to-left would provide us with a distribution of
erasures having an appropriate density per window to be
recovered. Moreover, suppose that the code properties are such that
inversion in the decoding direction is possible. Then, we would
possibly increase the decoding capability leading to less information
loss.

In order that such a scenario can work we should be able to compute 
a guard space (a sufficient large sequence of symbols without erasures).
We will explain in Section~\ref{Stronger} how this can be achieved.

%Convolutional codes consider the information as a complete
%sequence where $z$ indicates in which instant did each
%coefficient arrive. The sliding window property allows
%convolutional codes to move along the sequences and frame
%different sizes of window providing them with a lot of
%flexibility.  In order to apply a process where we can slide from
%right-to-left we will construct codes that are able to generate
%these sequences in inverted order, that is, from instant $l$ to
%$0$. Therefore, these codes will be able to execute a recovering
%process that starts at the end of the sequence and progresses up
%till the beginning. 
We will refer to the left-to-right decoding process 
as \textit{forward decoding} and to the  inverted (from right-to-left) 
recovering process as \textit{backward decoding}.

We will show how convolutional codes allow a ``forward and backward
flexibility'' which, together with extra algebraic properties imposed on the
codes, leads to the recovering of erasure patterns that block codes
cannot recover.  We recall the following results.

\begin{proposition}(\cite[Proposition 2.9]{hu08c}) Let
  $\mathcal{C}$ be an $(n,k,\delta)$ convolutional code with minimal generator
  matrix $G(z)$. Let $\overline{G}(z)$ be the matrix obtained by
  replacing each entry $g_{ij}(z)$ of $G(z)$ by
  $\overline{g_{ij}}(z):=z^{\delta_{j}}g_{ij}(z^{-1})$, where
  $\delta_j$ is the $j$-th column degree of $G(z)$. Then,
  $\overline{G}(z)$ is a minimal generator matrix of an
  $(n,k,\delta)$ convolutional code $\overline{\mathcal{C}}$, having the
  characterization
  \[
  \vect{v}_0 +\vect{v}_1 z + \cdots + \vect{v}_{s-1} z^{s-1} +
  \vect{v}_s z^{s} \in \mathcal{C}
  \]
  if and only if
  \[
  \vect{v}_s +\vect{v}_{s-1} z + \cdots + \vect{v}_{1} z^{s-1} +
  \vect{v}_0 z^{s} \in \overline{\mathcal{C}}.
  \]
\end{proposition}
We call $\overline{\mathcal{C}}$ the \textbf{reverse code} of
$\mathcal{C}$. Similarly, we denote by
$\overline{H}(z)=\sum_{i=0}^{\nu}\overline{H}_{i}z^{i}$ the
parity-check matrix of $\overline{\mathcal{C}}$. 

\begin{remark}
  Massey introduces in~\cite{ma64} the notion of reversible
  convolutional codes over the binary field. The definition has a
  natural generalization to the nonbinary situation. We would call a
  code reversible in the sense of Massey if
  $\overline{\mathcal{C}}=\mathcal{C}$ and where
  $\overline{\mathcal{C}}$ is the reverse code as defined above.
\end{remark}

Next we will use $\overline{\mathcal{C}}$ to explain the backward
decoding.  Although $\mathcal{C}$ and $\overline{\mathcal{C}}$ have
the same free distance $\dfree$, they may have different values for
the column distances, since the truncations of the code words
$\vec{v}(z)=\sum_{i=0}^{s}\vect{v}_i z^i$ and
$\overline{\vec{v}}(z)=\sum_{i=0}^{s}\vect{v}_{s-i}z^{i}$ do not
involve the same coefficients:
\begin{align*}
  d_{j}^{c}(\mathcal{C}) = & \min \left\{\wt(\vect{v}_{[0,j]}(z)) \ | \ \vec{v}(z) \in \mathcal{C}
    \quad \text{and} \quad \vect{v}_{0}\neq \vec{0} \right\}\\
  =  & \min \left\{\sum_{i=0}^{j}\wt(\vect{v}_{i}) \ | \ \vec{v}(z) \in \mathcal{C}
    \quad \text{and} \quad \vect{v}_{0}\neq \vec{0}
  \right\}\\
  d_{j}^{c}(\overline{\mathcal{C}}) = & \min \left\{\wt(\overline{\vect{v}}_{[0,j]}(z)) \ | \ \overline{\vec{v}}(z) \in \overline{\mathcal{C}}
    \quad \text{and} \quad \overline{\vect{v}}_{0}\neq \vec{0} \right\}\\
  = &  \min  \left\{
    % \small
    \sum_{i=0}^{j}\wt(\vect{v}_{s-i}) \ | \ \vec{v}(z) \in \mathcal{C}
    \  \text{and} \  \vect{v}_{s}\neq \vec{0} 
  \right\}.
\end{align*}
% \normalsize

Similar to the forward decoding process, in order to achieve
maximum recovering rate per window when recovering using backward
decoding, we need the column distances of $\overline{\mathcal{C}}$
to be maximal up to a point. This leads to the following
definition.

\begin{definition}
  Let $\mathcal{C}$ be an MDP $(n,k,\delta)$ convolutional code.
  We say that $\mathcal{C}$ is a reverse-MDP convolutional code
  if the reverse code $\overline{\mathcal{C}}$ of $\mathcal{C}$
  is an MDP code as well.
\end{definition}

As previously explained, reverse-MDP convolutional codes are
better candidates than MDP convolutional codes for recovering
over the erasure channel. In analogy to Corollary~\ref{main} we have 
the result:
\begin{theorem}                              \label{main-rev}
  Let $\mathcal{C}$ be an $(n,k,\delta)$ reverse-MDP
  convolutional code.  If in any sliding window of length $(L+1)n$ at
  most $(L+1)(n-k)$ erasures occur in a transmitted sequence, then we
  can completely recover the sequence in polynomial time in $\delta$
  by iteratively decoding the symbols `from right to left'.
\end{theorem}
The proof is completely analogous to the one given
in~Theorem~\ref{main2} and Corollary~\ref{main}.

The following theorem shows that the existence of this class of codes
is guaranteed over fields with enough number of elements.

\begin{theorem}\label{existence}
  Let $k$, $n$ and $\delta$ be positive integers. An
  $(n,k,\delta)$ reverse-MDP convolutional code exists over a
  sufficiently large field.
\end{theorem}

% Before proving Theorem \ref{existence} we would like to recall
% the main idea of the proof of the existence of MDP
% convolutional codes \cite{hu05} since the proof or our theorem
% will be described in a similar way.
% 
The set of all convolutional codes forms a quasi-projective
variety \cite{ha77p} that can be also seen as a Zariski
open subset of the projective variety described in
\cite{ra94,ro99a1}. In \cite{hu05},  it was shown that MDP
codes form a generic set when viewed as a subset of the
quasi-projective variety of all $(n,k,\delta)$ convolutional
codes. Following similar ideas to the ones in the proof of the
existence of MDP convolutional codes \cite{hu05},  we will show
that reverse-MDP codes form a nonempty Zariski open set of the
quasi-projective variety of generic convolutional codes and
moreover, that the components of the elements of this set are
contained in a finite field or a finite extension of it.

\begin{remark}
  The proof given in \cite{hu05} is based on a systems theory
  representation of convolutional codes $\mathcal{C}(A,B,C,D)$.
  Since this  is closely related to the submodule point of
  view that we consider and since a convolutional code can be
  represented in either way we will use in our proof the same
  notions as in \cite{hu05}. See \cite{ro96a1,ro99a,ro01} for
  further references and details on systems theory
  representations. In \cite{hu05},  the set of MDP convolutional
  codes is described by sets of matrices
  $\{F_{0},F_{1},\ldots,F_{L}\}$ that form a matrix
  $\mathcal{T}_{L}$ with the MDP property. The matrices
  $\{F_{0},F_{1},\ldots,F_{L}\}$ are directly related to the
  representation $(A,B,C,D)$ and have their  elements in
  $\bar{\mathbb{F}}$, the closure of a certain finite base field
  $\mathbb{F}$. $\bar{\mathbb{F}}$ is therefore an infinite field. Based on
  the minors that must be nonzero in
  $\mathcal{T}_{L}$ for $\mathcal{C}$ to be MDP,  a set of finitely
  many polynomial equations is obtained. This set describes the
  codes that do not satisfy the MDP property. The zeros of each of
  these polynomials describe a proper algebraic subset of
  $\bar{\mathbb{F}}^{(L+1)(n-k)k}$. The complement of these
  subsets are nonempty Zariski open sets in
  $\bar{\mathbb{F}}^{(L+1)(n-k)k}$. Their intersection is a
  nonempty Zariski open set since there is a finite number of
  them. Thus, the set of MDP codes forms a nonempty Zariski open
  subset of the quasi-projective variety of convolutional codes.
  \hfill$\square$
\end{remark}
Now we are ready to proof Theorem \ref{existence}.

\begin{IEEEproof}
  Let $\mathbb{F}$ be a finite field and $\bar{\mathbb{F}}$ be its
  algebraic closure. Following a similar reasoning to the one in
  \cite{hu05}, we will show that the set of reverse-MDP convolutional
  codes forms a generic set when viewed as a subset of the
  quasi-projective variety of all $(n,k,\delta)$ convolutional codes,
  by showing that it is the intersection of two nonempty Zariski open sets: the
  one of MDP codes and the one of the codes whose reverse is MDP.

  As shown in \cite{hu05}, there exist sets of finitely many
  polynomial equations whose zero sets describe those convolutional
  codes that are not MDP. Each of these sets is a proper subset of
  $\bar{\mathbb{F}}^{(L+1)(n-k)k}$, and its complement is a nonempty
  Zariski open set in $\bar{\mathbb{F}}^{(L+1)(n-k)k}$.  Let $\{
  W_j\}_{j=0}^{\theta}$, $\theta<\infty$, denote those complements.
  With a similar set of finitely many polynomial equations, one can
  describe those codes whose reverse ones are not MDP. These zero sets
  are proper algebraic sets over $\bar{\mathbb{F}}^{(L+1)(n-k)k}$, and
  the complement of those, let us denote them by $\{
  U_j\}_{j=0}^{\phi}$, $ \phi<\infty$, are also nonempty Zariski open
  sets in $\bar{\mathbb{F}}^{(L+1)(n-k)k}$.  Let $V$ be the
  intersection of all these sets  
  \[
  V= \left(\bigcap_{j=0}^{\theta}W_j \right)
  \bigcap\left(\bigcap_{j=0}^{\phi}U_j \right).
  \]
  Thus $V$ is a nonempty Zariski open set since there are finitely
  many sets in the intersection. $V$ describes the set of reverse-MDP
  codes. If we take one element in $V$, i.e., we select the matrices
  $\{F_0,F_1,\ldots F_L \}$ that represent a certain reverse-MDP code
  $\mathcal{C}$, then we have finitely many entries.  Either all of
  them belong to $\mathbb{F}$ or they all belong to a finite extension
  field of $\mathbb{F}$.  Choosing this extension, implies that we can
  always find a finite field where reverse-MDP codes exist.
\end{IEEEproof}

\begin{remark}
  The equations characterizing the set of reverse-MDP convolutional
  codes can be made very explicit for codes of degree $\delta$ where
  $(n-k)\mid\delta$. Let $H(z)=H_0+H_1 z+ \cdots+H_{\nu}z^{\nu}$ be a
  parity-check matrix of the code.  The reverse code has parity-check
  matrix $\overline{H}(z)=H_{\nu}+H_{\nu -1}z+\cdots + H_0 z^{\nu}$.
  Then $\overline{H}(z)$ gives a reverse-MDP code if and only if the
  algebraic conditions of an MDP code of Theorem \ref{mdp} hold for
  $\overline{H}(z)$, and, in addition, every full size minor of the
  matrix
\[
\begin{sbmatrix}{cccc}
H_{\nu} & H_{\nu-1} & \cdots & H_{\nu -L} \\
       & H_{\nu}    & \cdots & H_{\nu-L-1} \\
       &           & \ddots & \vdots \\
       &           &        & H_{\nu} 
\end{sbmatrix}
\]
formed from the columns with indices
$j_{1},j_{2},\ldots,j_{(L+1)(n-k)}$ having the property that $j_{s(n-k)+1}>sn$, for
$s=1,2,\ldots,L$, is nonzero.
\end{remark}

\begin{example}
  Let $\mathcal{C}$ be the $(2,1,1)$ convolutional code over
  $\mathbb{F}_{2^5}$ given by the parity-check matrix
  \[
  H(z)=\begin{sbmatrix}{cc} 1 + \alpha ^{25} z + \alpha^{5} z^{2} 
  & \alpha^{15} + \alpha^{10} z + \alpha^{3} z^{2}
  \end{sbmatrix}
  \]
  where $\alpha$ satisfies $\alpha^{5}+\alpha^{2}+1=0$. $\mathcal{C}$
  is an MDP code since in the matrix $\mathcal{H}_L$
  \[
  \mathcal{H}_L=
  \begin{sbmatrix}{cccccc}
    1           & \alpha^{15} & 0           & 0           & 0 & 0 \\
    \alpha^{25} & \alpha^{10} & 1           & \alpha^{15} & 0 & 0\\
    \alpha^{5}  & \alpha^{3}  & \alpha^{25} & \alpha^{10} & 1 & \alpha^{15} \\
  \end{sbmatrix},
  \]
  every $3\times 3$ non-trivially zero minor is nonzero.
  Moreover, the reverse code
  $\overline{\mathcal{C}}$ is defined by the matrix
  \begin{multline*}
  \overline{H}(z)=\begin{sbmatrix}{cc} \alpha^{5} + \alpha ^{25} z + z^{2} 
  & \alpha^{3} + \alpha^{10} z + \alpha^{15} z^{2}
  \end{sbmatrix}\\
  = \overline{H}_0+\overline{H}_1 z + \overline{H}_2 z^{2} .
  \end{multline*}
  $\overline{\mathcal{H}}_{L}$ is of the form
  \[
  \overline{\mathcal{H}}_L=
  \begin{sbmatrix}{cccccc}
    \alpha^{3}  & \alpha^{5}  & 0           & 0           & 0           & 0 \\
    \alpha^{10} & \alpha^{25} & \alpha^{3}  & \alpha^{5}  & 0           & 0 \\
    \alpha^{15} & 1           & \alpha^{10} & \alpha^{25} & \alpha^{3}  & \alpha^{5} \\
  \end{sbmatrix},
  \]
  for which again every $3\times 3$ non-trivially zero minor is nonzero.
  This shows that $\overline{\mathcal{C}}$ is an MDP code
  and therefore $\mathcal{C}$ is a reverse-MDP convolutional
  code.  \hfill$\square$
\end{example}

One can apply the backward process  to the received
sequence, taking into account that \small
\begin{align}\label{eq7}
  \begin{sbmatrix}{ccccccccc}
    \overline{H}_{\nu}  \!\!\!\!\! &  \overline{H}_{\nu-1}  \!\!\!\!\! & \cdots \!\!\!\!\! &  \overline{H}_{\nu-j}   \!\!\!\!\! & \cdots  \!\!\!\!\! &  \overline{H}_{0} \!\!\!\!\! \\
    \!\!\!\!\! &  \overline{H}_{\nu}    \!\!\!\!\! & \cdots \!\!\!\!\! &  \overline{H}_{\nu-j+1} \!\!\!\!\! & \cdots  \!\!\!\!\! &  \overline{H}_{1}   \!\!\!\!\! & \overline{H}_{0} \!\!\!\!\! \\
    \!\!\!\!\! &                        \!\!\!\!\! & \ddots \!\!\!\!\! &                         \!\!\!\!\! &         \!\!\!\!\! &                    \!\!\!\!\! &          \!\!\!\!\!  &  \ddots \!\!\!\!\! \\
    \!\!\!\!\! & \!\!\!\!\! & \!\!\!\!\! & \overline{H}_{\nu}
    \!\!\!\!\! & \cdots \!\!\!\!\! & \overline{H}_{j} \!\!\!\!\!
    & \overline{H}_{j-1} \!\!\!\!\! & \cdots \!\!\!\!\! &
    \overline{H}_{0}
  \end{sbmatrix}
  \begin{sbmatrix}{c}
    \!\!\! \vect{v}_{t+j} \!\!\! \\
    \!\!\! \vect{v}_{t+j-1} \!\!\! \\
    \!\!\! \vdots \!\!\! \\
    \!\!\! \vect{v}_{t+1} \!\!\! \\
    \!\!\! \vect{v}_t \!\!\!
  \end{sbmatrix}=\vec{0}
\end{align}
\normalsize if and only if
\begin{align}\label{eq8}
  \begin{sbmatrix}{ccccccccc}
    F_{\nu}  \!\!\!\!\! &  F_{\nu-1}  \!\!\!\!\! & \cdots   \!\!\!\!\! &  F_{\nu-j}   \!\!\!\!\! & \cdots  \!\!\!\!\! &  F_{0} \!\!\!\!\! \\
    \!\!\!\!\! &  F_{\nu}    \!\!\!\!\! & \cdots   \!\!\!\!\! &  F_{\nu-j+1} \!\!\!\!\! & \cdots  \!\!\!\!\! &  F_{1}  \!\!\!\!\! & F_{0} \!\!\!\!\! \\
    \!\!\!\!\! &             \!\!\!\!\! &  \ddots  \!\!\!\!\! &              \!\!\!\!\! &         \!\!\!\!\! &         \!\!\!\!\! &        \!\!\!\!\! &  \ddots  \!\!\!\!\! \\
    \!\!\!\!\! & \!\!\!\!\! & \!\!\!\!\! & F_{\nu} \!\!\!\!\! &
    \cdots \!\!\!\!\! & F_{j} \!\!\!\!\! & F_{j-1} \!\!\!\!\! &
    \cdots \!\!\!\!\! & F_{0}
  \end{sbmatrix}
  \begin{sbmatrix}{c}
    \!\!\! \overline{\vect{v}}_{t+j} \!\!\! \\
    \!\!\! \overline{\vect{v}}_{t+j-1} \!\!\! \\
    \!\!\! \vdots \!\!\! \\
    \!\!\! \overline{\vect{v}}_{t+1} \!\!\! \\
    \!\!\! \overline{\vect{v}}_t \!\!\!
  \end{sbmatrix}=\vec{0}, 
\end{align}
where $F_i=\mathcal{J}_{(n-k)}\overline{H}_i\mathcal{J}_n$,  for
$i=0,1,\ldots,\nu$,
$\overline{\vect{v}}_{j}=\vect{v}_{j}\mathcal{J}_{n}$,  for
$j=0,1,2,\ldots$,  and $\mathcal{J}_{r}$ is an $r\times r$ matrix
of the form
\[
\mathcal{J}_{r}=
\begin{sbmatrix}{ccccc}
  0      & 0 & \cdots & 0 & 1 \\
  0      & 0 & \cdots & 1 & 0 \\
  \vdots &   & %\iddots 
&  & \vdots \\
  0      & 1 & \cdots & 0 & 0 \\
  1      & 0 & \cdots & 0 & 0 \\
\end{sbmatrix}
\quad \text{for} \quad r=1,2,\ldots,n.
\]

Since both matrices are related by permutations of rows and
columns, then the matrix in expression (\ref{eq7}) satisfies the
MDP property if and only if the matrix in expression (\ref{eq8})
does. So one can work with the latter to recover the erasures
without the need of any  transformation on the received sequence.

We revisit the situation of Example \ref{ex4} and show how reverse-MDP
convolutional codes can recover the erasures that were rendered
undecodable by MDP convolutional codes.\\ 
\textbf{Example \ref{ex4}(cont).}
Assume that the $(2,1,50)$ code of Example \ref{ex4} 
is a reverse-MDP convolutional code. The reverse code
$\overline{\mathcal{C}}$ has the same recovering rate per window as
$\mathcal{C}$.

Recall that we were not able to recover the received sequence using a
left-to-right process.  We will do this by using a backward
recovering.

Once we have received $100$ symbols of $C$ we can recover part of the
past erasures.  If we take the following window
\[
\overbrace{\vect{v}\vect{v}\star\star\vect{v}
\vect{v}\star\star\ldots\vect{v}\vect{v}\star\star}^{(B)180}|
\overbrace{\vect{v}\vect{v}\ldots\vect{v}}^{(C)100}
\]
and use the reverse code $\overline{\mathcal{C}}$ to solve the
inverted system, then we can recover the erasures in $B$.  Moreover,
taking $100$ correct symbols from $G$, the $60$ erasures in $F$ and $60$
more correct symbols from $E$
\[
\overbrace{\vect{v}\vect{v}\ldots
  \vect{v}}^{(E)60}\overbrace{\star\star\ldots
  \star}^{(F)60}|\overbrace{\vect{v}\vect{v}\ldots
  \vect{v}}^{(G)100}
\]
we can in the same way recover block $F$.  We thus recovered
$150$ erasures which is more than $59$\% of the erasures that occurred
in that part of the sequence.  \hfill$\square$ \medskip

In the previous example we showed how reverse-MDP convolutional
codes and the backward process make it possible to recover
information that would already be considered as lost by an MDS
block code, or by an MDP convolutional code. We use a portion of the
guard space not only to possibly recover the next burst of erasures, but
additionally, to recover previous ones. We can do this as soon
as we receive enough correct symbols;  we do not need to wait
until we receive a whole new block.

If we would allow this backward process to be complete, that is,
to go from the end of the sequence up to the beginning, we would
recover a lot  more information. We do not consider
this situation since it would imply that we need to wait until
the whole sequence was received in order to start recovering
right-to-left and that would not give better results than the
retransmission of lost packets.

The following Algorithm  presents the recovering algorithm for a
sequence of length $l$. The value $0$ represents a packet that has not
been received; $1$ represents a correctly received packet; $\vec{1}$,
a vector of ones, represents a guard space;
$\mathsf{findzeros}(\vect{v})$ is a function that returns a vector
with the positions of the zeros in $\vect{v}$, and
$\mathsf{forward}(\mathcal{C},j,\vect{v})$ and
$\mathsf{backward}(\overline{\mathcal{C}},j,\vect{v})$ are the forward
and backward recovering functions, respectively. They use the parity
check matrices of $\mathcal{C}$ and $\overline{\mathcal{C}}$ to
recover the erasures that happen in $\vect{v}$ within a window of size
$(j+1)n$.\clearpage

\noindent {\bf RECOVERING ALGORITHM} \\
%\small
{\bf Data:} $[\vect{v}_{0},\vect{v}_{1},\ldots,\vect{v}_{l}]$, the received sequence.\\
{\bf Result:} $[\vect{v}_{0},\vect{v}_{1},\ldots,\vect{v}_{l}]$, the corrected sequence.\\
\begin{algorithmic}[1]
  \STATE $i=0$
  \WHILE{$i\leq l$}
    \STATE $forwardsucces=0$
    \STATE $backwardsucces=0$
    \IF{$v_i=0$}
      \IF{$[v_{(i-\nu n)},\ldots,v_{i-1}]=\vec{1}$}
        \STATE $j=L$
        \WHILE{$forwardsucces=0$ {\bf and} $j\geq 0$}
          \IF{$\length(\mathsf{findzeros}([v_{i},\ldots,v_{i+(j+1)n-1}]))$ \\
              $\leq (j+1)(n-k)$}
            \STATE $[v_{i-\nu n},\ldots,v_{i+(j+1)n-1}]$\\
            $= \mathsf{forward}(\mathcal{C},j,[v_{i-\nu n},\ldots,v_{i+(j+1)n-1}])$
            \STATE $forwardsucces=1$
            \STATE $i=i+(j+1)n-1$
          \ENDIF
          \STATE $j=j-1$
        \ENDWHILE
        \IF{$forwardsucces\neq 1$}
          \STATE $aux=\mathsf{findzeros}([v_{i},\ldots,v_{i+(L+1)n-1}])$
          \STATE $k=i+aux[\length(aux)]-1$
          \WHILE {$backwardsucces=0$ {\bf and} $k\leq l$}
            \IF{$[v_{k},\ldots,v_{k+\nu n-1}]=\vec{1}$}
              \STATE $j=L$
              \WHILE{$backwardsucces=0$ {\bf and} $j\geq 0$}
                \IF{$\length(\mathsf{findzeros}([v_{k-(j+1)n},
                              \ldots,v_{k-1}]))$
                              $\leq(j+1)(n-k)$}
                  \STATE $[v_{k-(j+1)n},\ldots,v_{k+\nu n-1}]=$\\
                  $\mathsf{backward}(\overline{\mathcal{C}},j,$
                  $[v_{k-(j+1)n},\ldots,v_{k+\nu n-1}])$
                  \STATE $backwardsucces=1$
                  \STATE $i=k+\nu n-1$
                \ENDIF
                \STATE $j=j-1$
              \ENDWHILE
            \ENDIF
            \STATE $k=k+1$
          \ENDWHILE
        \ENDIF
      \ENDIF
      \STATE $i=i+1$
    \ENDIF
  \ENDWHILE
\normalsize
\end{algorithmic}

The algorithm works as follows: It starts moving forward
(left-to-right) along the received sequence. Once a first erasure is
found, it checks if there is enough guard space
previous to the erasure.
	%Forward If enough clean memory 
If this occurs, it takes the next window of length $(j+1)n$ and checks
if the number of erasures is not greater than $(j+1)(n-k)$.
%				We can recover
If this condition holds, the recovery process is successful, that is,
our system has a unique solution, and we can move on to the next
window and start the process again.
				
%				We can't recover
On the other hand, if there are too many erasures, the window size
will be decreased until finding an erasure rate that can be recovered.
If no smaller window size is suitable for a successful recovery, the
backward process will start from the end of this window.
%		    Backward 
Since now we move right-to-left, the algorithm tests if there exists a
guard space after this window.
%		     If enough clean memory
In case this is true, the next step is to check if the erasure rate
moving to the left along the sequence allows the recovery. As in the
forward process, when the system cannot be solved, the window size
will decrease to a size where the number of erasures does not surpass
$(j+1)(n-k)$. If a window with these characteristics is found, this
part of the sequence will be recovered and the forward recovering
process will be retaken from this point on.  In case such window does
not exist, that part of the sequence will be considered as not
possible to be recovered and the forward recovering process will
restart at this point.
         
%					We can recover
%					 We can't recover
%				 If not enough clean memory

\begin{remark}
  Note  that the first and the last blocks of length $(j+1)n$ of
  the sequence (when using $\mathcal{C}$ and
  $\overline{\mathcal{C}}$, respectively) do not need the use of
  previous guard space since we assume that $\vect{v}_{i}=0$, for
  $i < 0$ and $i > l$, which allows us to solve the following systems
  \[
  \begin{sbmatrix}{cccc}
    H_0 \\
    H_1 & H_0 \\
    \vdots & \vdots & \ddots \\
    H_j & H_{j-1} & \cdots & H_0
  \end{sbmatrix}
  \begin{sbmatrix}{c}
    \vect{v}_0 \\
    \vect{v}_1 \\
    \vdots \\
    \vect{v}_j
  \end{sbmatrix}=\vec{0}, \quad j=0,1,\ldots,L,
  \]
  \[
  \begin{sbmatrix}{cccc}
    \!\! H_L & \!\!\! \cdots & \!\!\! H_{L-j+1}   & \!\!\! H_{L-j}     \!\! \\
    \!\!     & \!\!\! \ddots & \!\!\! \vdots  & \!\!\! \vdots  \!\! \\
    \!\!     & \!\!\!        & \!\!\! H_L     & \!\!\! H_{L-1}     \!\! \\
    \!\!     & \!\!\!        & \!\!\!         & \!\!\! H_{L}     \!\! \\
  \end{sbmatrix}
  \!\!
  \begin{sbmatrix}{c}
    \!\!\! \vect{v}_{l-j} \!\!\! \\
    \!\! \! \vect{v}_{l-j-1} \!\!\! \\
    \!\! \! \vdots \!\!\! \\
    \!\! \! \vect{v}_l \!\!\!
  \end{sbmatrix}=\vec{0}, \ j=0,1,\ldots,L.
  \]
  \hfill$\square$
\end{remark}

\section{Construction of reverse-MDP Convolutional   Codes}\label{ConstRev}

As we showed previously, reverse-MDP convolutional codes exist over
sufficiently large fields giving a good performance when decoding over
the erasure channel. Unfortunately, we do not have a general
construction for this type of codes because, for certain values of the
parameters, we do not know what is the relation between matrices
$H_{i}$ and matrices $\overline{H}_{i}$, $i=0,1,\ldots,\nu$. In this
section, we construct reverse-MDP codes for the case when $(n-k) \mid
\delta$ and $k>\delta$ ---situation in which we would need to give a
parity-check matrix--- or $k \mid \delta$ and $(n-k) > \delta$
---situation in which we would need to give a generator matrix.

Since reverse-MDP codes are codes satisfying both the forward and
the backward MDP property, we could try to modify
MDP convolutional codes such that the corresponding reverse codes are also MDP. 
Recall from \cite{hu08}  that in the construction of  MDP convolutional codes the following 
types of matrices play an essential role.
\begin{definition}\label{Toeplitz}
  Let $A$ be an $r \times r$ lower triangular Toeplitz matrix
  \[
  A=\begin{sbmatrix}{cccc}
    a_0    & 0      & \cdots & 0 \\
    a_1    & a_0    & \ddots & \vdots \\
    \vdots & \ddots & \ddots & 0 \\
    a_r & \cdots & a_1 & a_0
  \end{sbmatrix}.
  \]
  Let $s \in \{1,2,\ldots,r\}$. Suppose that $I:=\{ i_1,\ldots,
  i_s\}$ is a set of row indices of $A$, $J:=\{ j_1,\ldots,
  j_s\}$ is a set of column indices of $A$, and that the elements
  of each set are ordered from smallest to largest. We denote by
  $A_{J}^{I}$ the sub-matrix of $A$ formed by intersecting the
  columns indexed by the members of $J$ and the rows indexed by
  the members of $I$. A sub-matrix of $A$ is said to be
  \textbf{proper} if, for each $t\in\{1,2,\ldots,s\}$, the
  inequality $j_t \leq i_t$ holds. The matrix $A$ is said to be
  \textbf{superregular} if every proper sub-matrix of $A$ has a
  nonzero determinant.
\end{definition}
% To make it clear to the reader, the proper sub-matrices of $A$ are
% those which don't have too many zeros, that is, those which have
% the chance to have a non zero determinant.  We will call the
% minors corresponding to those matrices \textit{non trivially
% zero} minors.

\begin{remark}
  In the case $A$ is not a lower triangular matrix, but a lower block
  triangular matrix of size $\gamma(n-k)\times \gamma k$, where each
  block has size $(n-k)\times k$, a proper sub-matrix of $A$ is a
  sub-matrix $A^{I}_{J}$ such that the inequality $j_t \leq \left\lceil
    \frac{i_t}{n-k} \right\rceil k$ holds, for each
  $t\in\{1,2,\ldots,\min\{\gamma(n-k),\gamma k \} \}$.
\end{remark}

In \cite{hu08}, the parity-check matrix of an MDP convolutional code
was constructed using its systematic form, that is,
$\hat{\mathcal{H}}_{L}=[I_{(L+1)(n-k)} ~|~\hat{H}]$, where
$I_{(L+1)(n-k)}$ is the identity matrix of size $(L+1)(n-k)$ and
$\hat{H}$ is a $(L+1)(n-k) \times (L+1)k$ lower block triangular
superregular matrix. After left multiplication by an invertible matrix
and a suitable column permutation on the systematic expression
$\hat{\mathcal{H}}_{L}$ we can obtain the parity-check matrix
$\mathcal{H}_{L}$ given in (\ref{eq5}).  Note that the nonzero minors
of any size of the lower block triangular superregular matrix
$\hat{H}$ translate into nonzero full size minors of
$\mathcal{H}_{L}$, property that characterizes MDP convolutional
codes.

Motivated by this idea we introduce the following
matrices. 
%now propose a construction of reverse-MDP
%convolutional codes based on the use of specific matrices that 
%are superregular forward and backward.  The minors of these
%matrices will translate into full size minors of the parity-check
%matrices of $\mathcal{C}$ and $\overline{\mathcal{C}}$.

\begin{definition}
  We say a superregular matrix $A$ is
  \textbf{reverse-superregular} if the matrix
  \[
  A_{\rev}=\begin{sbmatrix}{cccc}
    a_r    & 0      & \cdots & 0 \\
    a_{r-1} & a_r    & \ddots & \vdots \\
    \vdots & \ddots & \ddots & 0 \\
    a_0 & \cdots & a_{r-1} & a_r
  \end{sbmatrix}
  \]
  is superregular.
\end{definition}

  These matrices may be hard to find since not every superregular
  matrix is a reverse-superregular matrix.
  
  \begin{example} Let   
  \[
  A=\begin{sbmatrix}{cccc}
    1        & 0        & 0      & 0 \\
    \alpha   &       1  & 0      & 0 \\
    \alpha^3 & \alpha   & 1      & 0 \\
    \alpha & \alpha^3 & \alpha & 1
  \end{sbmatrix},
  \]
  where $\alpha^3 + \alpha + 1=0$.  We can easily check that $A$  is superregular over
  $\mathbb{F}_{8}$.  However, its reverse matrix
  \[
  A_{\rev}=\begin{sbmatrix}{cccc}
    \alpha   & 0        & 0        & 0 \\
    \alpha^3 & \alpha   & 0        & 0 \\
    \alpha   & \alpha^3 & \alpha   & 0 \\
    1 & \alpha & \alpha^3 & \alpha
  \end{sbmatrix}
  \]
  is not superregular since $\left\vert \begin{array}{ccc}
      \alpha^3 & \alpha & 0        \\
      \alpha   & \alpha^3 & \alpha \\
      1 & \alpha & \alpha^3
    \end{array} \right\vert =0$. 
    \hfill$\square$
\end{example}

  One can think that only superregular matrices that are
  symmetric with respect to the  lower diagonal can be
  reverse-superregular. 

  \begin{example} Let \[ B=
  \begin{sbmatrix}{cccc}
    1      & 0      & 0      & 0 \\
    \alpha & 1      & 0      & 0 \\
    \alpha & \alpha & 1      & 0 \\
    1 & \alpha & \alpha & 1
  \end{sbmatrix}.
  \]
  Then $B$ is a reverse-superregular matrix over $\mathbb{F}_8$, where
  $\alpha^3 + \alpha^2 + 1=0$, because $B=B_{\rev}$ and the minor
  property holds.
  \end{example}

  The following example shows that, in fact, superregular matrices that are not
  symmetric with respect to the lower diagonal can be
  reverse-superregular as well.
  \begin{example}
 Let  $C$ be the matrix below and $C_{\rev}$ its corresponding reverse matrix
  \[
  C=\begin{sbmatrix}{cccc}
    \! 1        & 0        & 0        & 0 \! \\
    \! \alpha^4 & 1        & 0        & 0 \! \\
    \! \alpha^6 & \alpha^4 & 1        & 0 \! \\
    \! \alpha^3 & \alpha^6 & \alpha^4 & 1 \!
  \end{sbmatrix} , \quad C_{\rev}=\begin{sbmatrix}{cccc}
    \! \alpha^3 & 0        & 0        & 0 \! \\
    \! \alpha^6 & \alpha^3 & 0        & 0 \! \\
    \! \alpha^4 & \alpha^6 & \alpha^3 & 0 \! \\
    \! 1 & \alpha^4 & \alpha^6 & \alpha^3 \!
  \end{sbmatrix}.
  \]
  Both $C$ and $C_{\rev}$ are superregular matrices over
  $\mathbb{F}_{8}$ implying that $C$ is a reverse-superregular matrix.
  \hfill$\square$
\end{example}

\medskip

Due to the importance that these matrices have in our
construction, in the following subsection we present several
tools to generate them.

\subsection{Construction of reverse-superregular
  matrices}\label{ConstMat}

Superregular matrices have been previously studied in relation to MDP
convolutional codes.  Minimum required field size necessary for
constructing an MDP code and a study of matrix or code transformations
that preserve superregularity can be found in the literature (see
\cite{hu05,hu08c,hu08,ke06}). In \cite{gl06} a concrete construction
of MDP codes is given, although over a field of size much larger than
the minimum possible for those parameters. In \cite{gl06} it was
conjectured that for every $l \geq 5$ one can find a superregular $l
\times l$-Toeplitz matrix over $\mathbb{F}_{2^{l-2}}$. This remained
an open question.

In this section we give a method of obtaining reverse-superre\-gu\-lar
matrices over fields of characteristic $p$ that requires less time
than an exhaustive computer search. We also present matrix
transformations that preserve the reverse-superregular pro\-per\-ty.
Although we cannot specify the minimum field size required for given
parameters, we can ensure that the matrices obtained with this method
are reverse-superregular.  Since reverse-superregularity is a more
restrictive condition than superregularity, it is reasonable to expect
that the field size needed to generate an $l\times
l$-reverse-superregular Toeplitz matrix over fields of characteristic
$2$ is larger than that for general superregularity.  In this
construction, the size of the field will be $\mathbb{F}_{2^{l-1}}$
which is larger than the size conjectured in \cite{gl06} for general
superregularity.
\begin{theorem}\label{const}
  Let $p(x)$ be an irreducible polynomial of degree $n$ over
  $\mathbb{F}_{p^n}$ and let $\alpha$ be a root, $p(\alpha)=0$.  
  Let $a(z)=\prod _{i=0}^{l-1}(1 + \alpha^i z)=a_0 +a_1 z + \ldots
  + a_l z^l$. If the matrix
  \[
  A=\begin{sbmatrix}{cccc}
    a_0    & 0      & \ldots & 0 \\
    a_1    & a_0    & \ddots & 0 \\
    \vdots & \ddots & \ddots & \vdots \\
    a_l & \ldots & a_1 & a_0
  \end{sbmatrix}
  \]
  is superregular, then the reversed matrix $A_{\rev}$ is
  superregular.
\end{theorem}
\begin{IEEEproof}
  By construction we have that
  \[
  a_{0}=1, \ a_{1}= \sum_{j=0}^{l-1}\alpha^{j}, \
  a_{2}=\sum_{j\neq k}\alpha^{j} \alpha^{k},
  \]
  \[
  \ a_{3}=\sum_{j\neq k \neq h} \alpha^{j}\alpha^{k}\alpha^{h}, \
  \ldots \ a_{l}=\prod_{j=0}^{l-1}\alpha^{j}.
  \]
  The minors of $A$ and $A_{\rev}$ can be described as polynomials in
  $\alpha$.  The following connection between these minors holds. If
  $\overline{\rho}(\alpha)$ denotes a minors of $A_{\rev}$ based on a
  set $I$ and $J$ of row and column indices, then there exists an
  integer $i$ such that
  \[
  \overline{\rho}(\alpha)=\alpha^{i}\rho(\frac{1}{\alpha}),
  \] where  $\rho(\alpha)$ is the minor of $A$ based on the same sets $I$ and $J$; 
 the power $i$ depends on the size of the minor. 
  The claim follows now. 
\end{IEEEproof}

Although we cannot decide \textit{a priori} which irreducible
polynomials will generate a reverse-superregular matrix, computer
search complexity is drastically reduced since the number of
irreducible polynomials generating a field is much smaller than the
field size. Search algorithms are efficient because only
superregularity needs to be tested since reverse-superregularity is
guaranteed by Theorem \ref{const}.

\medskip In the following we will present a few matrix transformations
that preserve reverse-superregularity. Note that the first two results
are the same as in \cite{hu08}, where several actions preserving
superregularity are studied.

\begin{theorem}
  Let $A$ defined as in Definition \ref{Toeplitz} be a
  reverse-superregular matrix over $\mathbb{F}_{p^{e}}$ and let
  $\alpha \in \mathbb{F}_{p^{e}}^{*}=\mathbb{F}_{p^{e}}\setminus
  \{0\}$.  Then
  \[
  \alpha \bullet A= \begin{sbmatrix}{ccccc}
    a_{0}           & 0                   & 0                   & \ldots & 0 \\
    \alpha a_{1}    & a_{0}               &  0                  & \ldots & 0 \\
    \alpha^{2}a_{2} & \alpha a_{1}        & a_{0}               &        & \vdots \\
    \vdots          & \vdots              &                     & \ddots & \vdots   \\
    \alpha^{l}a_{l} & \alpha^{l-1}a_{l-1} & \alpha^{l-2}a_{l-2} & \ldots & a_{0} \\
  \end{sbmatrix}
  \]
  is a reverse-superregular matrix.
\end{theorem}
\begin{IEEEproof}
  Since the minors in matrix $A$ are only transformed by factors
  of the form $\alpha^{i}$ in matrix $\alpha\bullet A$ and the
  same occurs for the minors of $(\alpha\bullet A)_{\rev}$, then
  reverse-superregularity is preserved.
\end{IEEEproof}
\begin{theorem}
  Let $A$ defined as in  Definition \ref{Toeplitz} be a reverse-superregular matrix over
  $\mathbb{F}_{p^{e}}$ and let $i \in \mathbb{Z} / e \mathbb{Z}$. Then
  \[
  i \circ A=\begin{sbmatrix}{ccccc}
    a_{0}^{p^{i}}    & 0               & 0               & \ldots & 0 \\
    a_{1}^{p^{i}}    & a_{0}^{p^{i}}   & 0               & \ldots & 0 \\
    a_{2}^{p^{i}}    & a_{1}^{p^{i}}   & a_{0}^{p^{i}}   &        & \vdots \\
    \vdots          & \vdots         &                 & \ddots & \vdots   \\
    a_{l}^{p^{i}}    & a_{l-1}^{p^{i}} & a_{l-2}^{p^{i}} & \ldots & a_{0}^{p^{i}} \\
  \end{sbmatrix}
  \]
  is a reverse-superregular matrix.
\end{theorem}
\begin{IEEEproof}
  In this case, the a minor of $i \circ A$ is
  the corresponding minor of $A$ to the power of $p^{i}$. The same occurs for $(i
  \circ A)_{\rev}$ and $A_{\rev}$, so we still have
  reverse-superregularity.
\end{IEEEproof}

The next theorem refers to the 
construction of Theorem \ref{const}.
\begin{theorem}\label{Thinverse}
  Let $A$ and $p(x)$ be as in the construction given in Theorem
  \ref{const}.  Then the matrix
  \[
  S=\begin{sbmatrix}{cccc}
    s_0    & 0      & \ldots & 0 \\
    s_1    & s_0    & \ddots & 0 \\
    \vdots & \ddots & \ddots & \vdots \\
    s_l & \ldots & s_1 & s_0
  \end{sbmatrix}
  \]
  where $s(z)=\prod_{i=0}^{l-1}(1 + (\alpha^{-1})^i z)=s_0 +s_1z
  + \ldots + s_l z^l$, is a reverse-superregular matrix.
\end{theorem}
\begin{IEEEproof}
  This is due to the fact that any minor of $A$, $\rho(\alpha)$,
  and the corresponding minor of $S$, $\sigma(\alpha)$, satisfy
  $\rho(\alpha)=\sigma(\frac{1}{\alpha})$. The same relation is
  given for the reversed matrices and therefore
  reverse-superregularity holds.
\end{IEEEproof}
Since the reciprocal polynomial $q(x)=x^{l}p(x^{-1})$ of an
irreducible polynomial $p(x)$ is irreducible too and the roots of
$q(x)$ are the inverse of the roots of $p(x)$, Theorem
\ref{Thinverse} reduces by half the number of irreducible
polynomials one must check since we can assume the same
behavior for $p(x)$ and $q(x)$. In this way computer searches
become again more efficient.

Note that not all actions preserving superregularity
preserve reverse-superregularity, as we show next.
 It is known \cite{hu08} that the inverse of a superregular matrix is a
  superregular matrix. The same does not occur for 
  reverse-superregularity since the reversed matrix of the
  inverse is not necessarily superregular.
\begin{example}
  The following $5\times 5$ matrix is reverse-superregular over
  $\mathbb{F}_{16}$ with $1+\alpha+\alpha^{4}=0$,
  \[
  Y=\begin{sbmatrix}{ccccc}
    1 & 0 & 0 & 0 & 0 \\
    \alpha^{12} & 1 & 0 & 0 & 0 \\
    \alpha^{4} & \alpha^{12} & 1 & 0 & 0 \\
    1 & \alpha^{4} & \alpha^{12} & 1 & 0 \\
    \alpha^{6} & 1 & \alpha^{4} & \alpha^{12} & 1
  \end{sbmatrix}.
  \]
  However, its inverse is not a reverse-superregular matrix since
  in
  \[
  (Y^{-1})_{\rev}=\begin{sbmatrix}{ccccc}
    \alpha^{14} & 0 & 0 & 0 & 0 \\
    \alpha^{13} & \alpha^{14} & 0 & 0 & 0 \\
    \alpha^{14} & \alpha^{13} & \alpha^{14} & 0 & 0 \\
    \alpha^{12} & \alpha^{14} & \alpha^{13} & \alpha^{14} & 0 \\
    1 & \alpha^{12} & \alpha^{14} & \alpha^{13} & \alpha^{14}
  \end{sbmatrix},
  \]
  the following minor is zero
  \[
  \left\vert
    \begin{array}{cccc}
      \alpha^{13} & \alpha^{14} & 0 & 0 \\
      \alpha^{14} & \alpha^{13} & \alpha^{14} & 0 \\
      \alpha^{12} & \alpha^{14} & \alpha^{13} & \alpha^{14} \\
      1 & \alpha^{12} & \alpha^{14} & \alpha^{13} 
    \end{array} \right\vert =0.
  \]
  \hfill$\square$
\end{example}

\bigskip

Once we have generated the necessary tools,  we can proceed to
construct reverse-MDP  codes.  

Let $(n-k)\mid \delta$ and $k>\delta$. We will 
extract appropriate columns and rows from a
reverse-superregular matrix to obtain a parity-check matrix
of a reverse-MDP code $\mathcal{C}$.% $\mathcal{C}$ and
%$\overline{\mathcal{C}}$ satisfy the MDP property.

\begin{theorem}\label{Thextract}
  Let $A$ be an $r \times r$ reverse-superregular matrix with
  $r=(L+1)(2n-k-1)$.  For $j=0,1,\ldots,L$,  let $I_j$  and $J_j$ be the following sets: 
    \begin{align*}
    I_j= & \left\{ (j+1)n + j(n-k-1),  \right. \\
    & \left. (j+1)n+ j(n-k-1)+1,\ldots,(j+1)(2n-k-1) \right\}, \\
 % \end{align*}
  %\begin{align*}
    J_j= & \left\{jn+j(n-k-1)+1, \right. \\
    & \left. jn+j(n-k-1)+2,\ldots,(j+1)n +j(n-k-1)\right\}, 
  \end{align*}
  and  $I$ and $J$ be the union of these sets
  \[
  I=\bigcup_{j=0}^{L}I_{j}, \quad \quad J=\bigcup_{j=0}^{L}J_{j}.
  \]
  Let $\widetilde{A}$ be the $(L+1)(n-k)\times (L+1)n$ lower
  block triangular sub-matrix with rows indexed by $I$ and columns indexed by $J$, i.e., 
  \[
  \widetilde{A}= A^{I}_{J}.
  \]
  Then every $(L+1)(n-k)\times (L+1)(n-k)$ full size minor of
  $\widetilde{A}$ formed from the columns with indices $1 \leq
  i_1<\cdots<i_{(L+1)(n-k)}$,  where $i_{s(n-k)}\leq sn$,  for
  $s=1,2,\ldots,L$, is nonzero.  \\  
  Moreover, the same property holds
  for $\widetilde{A}_{\rev}$.
\end{theorem}

We will use the above theorem to construct the lower block triangular
matrix $\mathcal{H}_{L}$. In $\mathcal{H}_{L}$ only matrices $H_{i}$
with $i\leq L$ appear. However, we know that
$H(z)=\sum_{i=0}^{\nu}H_{i}z^{i}$.  The condition $(n-k)\mid \delta$
and $k>\delta$ ensures that $L=\nu=\frac{\delta}{(n-k)}$ and
therefore, $H_{\nu}=H_{L}$. Then, all the matrices of the expansion of
$H(z)$ appear in $\mathcal{H}_L$ and we can describe $H(z)$ since the
blocks of the matrix $\widetilde{A}$ obtained in Theorem
\ref{Thextract} represent the matrices $H_i$.

Moreover, let $\mu_{j}$ be the maximum degree of all polynomials in
the $j$-th row of $H(z)$ and let $H_{\infty}$ be the matrix whose
$j$-th row is formed by the coefficients of $z^{\mu_{j}}$ in the
$j$-th row of $H(z)$.  In general $H_{\infty}\neq H_{\nu}$, but since
$(n-k)\mid \delta$, $H_{\nu}$ has full rank and so the two matrices
must coincide.  We have $\overline{H}_i=H_{\nu-i}$, for
$i=0,1,\ldots,\nu$, which yields that $\overline{H}(z)=H_{\nu} +
H_{\nu-1}z + \cdots + H_1 z^{\nu-1} +H_0 z^{\nu}$ is a parity-check
matrix of $\overline{\mathcal{C}}$. We can construct the lower block
triangular matrix $\overline{\mathcal{H}}_{L}$ using
$\widetilde{A}_{\rev}$, where the blocks of $\widetilde{A}_{\rev}$
represent the matrices $\overline{H}_{i}$. Then, we can describe
$\overline{H}(z)$. One can obtain $\overline{\mathcal{H}}_{L}$
inverting the positions of the blocks in the matrix $\mathcal{H}_{L}$
constructed with help of Theorem~\ref{Thextract} as well.

We illustrate the process with some examples.

\begin{example}
  In this example, we construct the parity-check matrix of a $(3,2,1)$
  reverse-MDP convolutional code $\mathcal{C}$ over $\mathbb{F}_{32}$
  using a $6 \times 6$ reverse-superregular matrix.  Let $\mu\in
  \mathbb{F}_{32}$ such that $\mu^5 +\mu^2 + 1=0$ and let $P$ be a
  reverse-superregular matrix constructed from $\mu$ over
  $\mathbb{F}_{32}$ as in Theorem \ref{const},
  \[
  P=\begin{sbmatrix}{cccccc}
    1          & 0           & 0           & 0          & 0          & 0 \\
    \mu^{15} & 1           & 0          & 0          & 0          & 0 \\
    \mu^{21} & \mu^{15} & 1           & 0          & 0          & 0 \\
    \mu^{23} & \mu^{21} & \mu^{15} & 1          & 0           & 0 \\
    \mu^{21} & \mu^{23} & \mu^{21} & \mu^{15} & 1           & 0 \\
    \mu^{10} & \mu^{21} & \mu^{23} & \mu^{21} & \mu^{15} & 1
  \end{sbmatrix}.
  \]
  According to the choice of sets $I$ and $J$ in Theorem
  \ref{Thextract} we obtain the matrix
  \[
  \mathcal{H}_L=\begin{sbmatrix}{cc}
    H_0 & O \\
    H_1 & H_0
  \end{sbmatrix}=
  \begin{sbmatrix}{cccccc}
    \mu^{21} & \mu^{15} & 1           & 0          & 0          & 0 \\
    \mu^{10} & \mu^{21} & \mu^{23} & \mu^{21} & \mu^{15} & 1
  \end{sbmatrix}, 
  \]
 leading to the parity-check matrix of $\mathcal{C}$ 
  \[H(z)=\begin{sbmatrix}{ccc} \mu^{21}+\mu^{10} z &
    \mu^{15}+\mu^{21} z & 1+\mu^{23} z
  \end{sbmatrix}.
  \]

  The parity-check matrix  $\overline{H}(z)$ for  $\mathcal{\overline{C}}$,  which is  given by 
  $\overline{H}(z)=\sum_{i=0}^{1}\overline{H}_{i}z^{i}=\sum_{i=0}^{1}H_{1-i}z^{i}$, 
  is  now  
  \[
  \overline{H}(z)=\begin{sbmatrix}{ccc} \mu^{10}+ \mu^{21} z &
    \mu^{21}+\mu^{15} z & \mu^{23}+ z
  \end{sbmatrix}.
  \]
 The matrix 
  \[
  \overline{H}_L=\begin{sbmatrix}{cccccc}
    \mu^{10}  & \mu^{21} & \mu^{23} & 0           & 0          & 0 \\
    \mu^{21} & \mu^{15} & 1 & \mu^{10} & \mu^{21} & \mu^{23}
  \end{sbmatrix}
  \]
  has equivalent properties to the ones of the matrix
  \[
  \mathcal{F}_L=\begin{sbmatrix}{cccccc}
    \mu^{23} & \mu^{21} & \mu^{10}   & 0          & 0          & 0 \\
    1 & \mu^{15} & \mu^{21} & \mu^{23} & \mu^{21} & \mu^{10}
  \end{sbmatrix},
  \]
  which we would have obtained applying Theorem \ref{Thextract}
  to the matrix $P_{\rev}$.  \hfill$\square$
\end{example}
\begin{example}
  We can use a $8 \times 8$ reverse-superregular matrix over
  $\mathbb{F}_{128}$ to construct a $(4,3,1)$ reverse-MDP
  convolutional code in the following way. Applying Theorem
  \ref{Thextract} to the matrix
  \[
  Q=\begin{sbmatrix}{cccccccc}
    1          & 0          & 0          & 0          & 0          & 0          & 0        & 0 \\
    \beta^{12} & 1          & 0          & 0          & 0          & 0          & 0        & 0\\
    \beta^{32} & \beta^{12} & 1          & 0          & 0          & 0          & 0        & 0 \\
    \beta^{45} & \beta^{32} & \beta^{12} & 1          & 0          & 0          & 0        & 0 \\
    \beta^{48} & \beta^{45} & \beta^{32} & \beta^{12} & 1          & 0          & 0        & 0 \\
    \beta^{41} & \beta^{48} & \beta^{45} & \beta^{32} & \beta^{12} & 1          & 0        & 0 \\
    \beta^{27} & \beta^{41} & \beta^{48} & \beta^{45} & \beta^{32} & \beta^{12} & 1        & 0 \\
    \beta^{21} & \beta^{27} & \beta^{41} & \beta^{48} &
    \beta^{45} & \beta^{32} & \beta^{12} & 1
  \end{sbmatrix}, 
  \]
  where $\beta^{7} + \beta^{6} + 1=0$, we obtain the matrix
  \begin{align*}
    \mathcal{H}_{L}=&
    \begin{sbmatrix}{cc}
      H_{0} & O \\
      H_{1} & H_{0}
    \end{sbmatrix} \\
    = & \begin{sbmatrix}{cccccccc}
      \beta^{45} & \beta^{32} & \beta^{12} &  1 & 0 & 0 & 0 & 0 \\
      \beta^{21} & \beta^{27} & \beta^{41} & \beta^{48} &
      \beta^{45} & \beta^{32} & \beta^{12} & 1
    \end{sbmatrix}.
  \end{align*}
  Then the parity-check matrix of $\mathcal{C}$ is \small
  \[
  H(z)=
  \begin{sbmatrix}{cccc}
    \beta^{45}+\beta^{21} z & \beta^{32} + \beta^{27} z &
    \beta^{12} + \beta^{41} z & 1 + \beta^{48}z
  \end{sbmatrix},
  \]
  \normalsize and the parity-check matrix of
  $\overline{\mathcal{C}}$ is \small
  \[
  \overline{H}(z)=
  \begin{sbmatrix}{cccc}
    \beta^{21} + \beta^{45} z & \beta^{27} + \beta^{32} z &
    \beta^{41} + \beta^{12} z & \beta^{48} + z
  \end{sbmatrix}.
  \]
  \normalsize \hfill$\square$
\end{example}
\medskip The same kind of construction can be applied in order to
obtain the generator matrix of a code. Transposing the
reverse-superregular matrix and adapting appropriately the sizes in
the row and column extraction, we obtain the following theorem similar
to Theorem \ref{Thextract}.
\begin{theorem}\label{Thextract2}
  Let $B$ be the transpose of an $r \times r$
  reverse-superregular matrix with $r=(L+1)(n+k-1)$.  For
  $j=0,1,\ldots,L$,  let $I_j$ and  $J_j$  as following \begin{align*}
    I_j= & \left\{ jn+j(k-1)+1, \right. \\
    & \left. jn+j(k-1)+2,\ldots,(j+1)n+j(k-1) \right\}, \\
 %\end{align*}
  %\begin{align*}
    J_j=& \left\{(j+1)n+j(k-1), \right. \\
    & \left. (j+1)n+j(k-1)+1,\ldots,(j+1)(n+k-1) \right\},
  \end{align*}
  and let $I$ and $J$ be the union of these sets
  \[
  I=\bigcup_{j=0}^{L} I_{j}, \quad \quad J=\bigcup_{j=0}^{L}
  J_{j}.
  \]
  Let $\widetilde{B}$ be the $(L+1)n\times (L+1)k$ upper block
  triangular sub-matrix with rows indexed by $I$ and columns indexed by $J$, i.e., 
  \[
  \widetilde{B}=B^{I}_{J}.
  \]
  Then every $(L+1)k\times (L+1)k$ full size minor of
  $\widetilde{B}$ formed from the columns with indices $1 \leq
  i_1<\cdots<i_{(L+1)k}$,  where $i_{sk+1}> sn$,  for $s=1,2,\ldots,L$,
  is nonzero.  \\ Moreover, the same property holds for
  $\widetilde{B}_{\rev}$.
\end{theorem}

In this case, the upper block triangular matrix $\widetilde{B}$ will
represent the matrix $\mathcal{G}_{L}$. In $\mathcal{G}_{L}$, only the
matrices $G_{i}$ with $i\leq L$ are involved. However,
$G(z)=\sum_{i=0}^{m}G_{i}z^{i}$.  When constructing generator
matrices, we need $k\mid \delta$ and $(n-k)>\delta$, so that
$L=m=\frac{\delta}{k}$ and $G_{\nu}=G_{L}$.  Then all the matrices in
the expansion of matrix $G(z)$ appear in $\mathcal{G}_{L}$ and we can
construct $G(z)$ using the blocks in $\widetilde{B}$ to describe the
matrices $G_{i}$.

Recall that $G_{\infty}$ is the matrix whose $j$-th column is formed
by the coefficients of $z^{\delta_{j}}$ in the $j$-th column of
$G(z)$, where $\delta_{j}$ is the $j$-th column degree of $G(z)$. As
in the parity-check matrix case, in general $G_{\infty}\neq G_m$, but
with $k \mid \delta$, $G_m$ has full rank and $G_{\infty}=G_m$. Now
$\overline{G}_i=G_{m-i}$ for $i=0,1,\ldots,m$ and the expression
$\overline{G}(z)=G_m + G_{m-1}z + \cdots + G_1 z^{m-1} +G_0 z^m$
describes a generator matrix of $\overline{\mathcal{C}}$.  The blocks
in matrix $\widetilde{B}_{\rev}$ represent now the matrices
$\overline{G}_{i}$ and can be used to construct $\overline{G}(z)$.
Since $\overline{G}_i=G_{m-i}$ for $i=0,1,\ldots,m$, one can use the
same blocks in $\mathcal{G}_{L}$ to construct $\overline{G}(z)$ as
well.

\begin{example}
  We  construct a  generator matrix of a $(3,1,1)$ code over
  $\mathbb{F}_{32}$. For this, we use the transpose of a $6\times
  6$ reverse-superregular matrix. Let
  $\gamma^{5}+\gamma^{4}+\gamma^{3}+\gamma^{2}+1=0$. We can apply
  Theorem \ref{Thextract2} to the matrix
  \[
  S=\begin{sbmatrix}{cccccc}
    1 & \gamma^{19} & \gamma^{16} & \gamma^{20} & \gamma^{5}  & \gamma^{16} \\
    0 & 1           & \gamma^{19} & \gamma^{16} & \gamma^{20} & \gamma^{5} \\
    0 & 0           & 1           & \gamma^{19} & \gamma^{16} & \gamma^{20} \\
    0 & 0           & 0           & 1           & \gamma^{19} & \gamma^{16} \\
    0 & 0           & 0           & 0           & 1           & \gamma^{19} \\
    0 & 0 & 0 & 0 & 0 & 1
  \end{sbmatrix},
  \]
  obtaining
  \[
  \mathcal{G}_{L}=\begin{sbmatrix}{cc}
    G_{0} & G_{1} \\
    O & G_{0}
  \end{sbmatrix}=
  \begin{sbmatrix}{cc}
    \gamma^{16} & \gamma^{16} \\
    \gamma^{19} & \gamma^{5} \\
    1           & \gamma^{20} \\
    0           & \gamma^{16} \\
    0           & \gamma^{19} \\
    0 & 1
  \end{sbmatrix}.
  \]
  The generator matrices of $\mathcal{C}$ and
  $\overline{\mathcal{C}}$ are
  \[
  G(z)=\begin{sbmatrix}{c}
    \gamma^{16}  + \gamma^{16} z \\
    \gamma^{19}  + \gamma^{5} z \\
    1            + \gamma^{20} z\\
  \end{sbmatrix}
  \quad \text{and} \quad \overline{G}(z)=\begin{sbmatrix}{c}
    \gamma^{16}  + \gamma^{16} z \\
    \gamma^{5}   + \gamma^{19} z \\
    \gamma^{20}  + z\\
  \end{sbmatrix}.
  \]
  \hfill$\square$
\end{example}

\section{Complete-MDP convolutional codes}
\label{Stronger}

We explained earlier how reverse-MDP codes can improve the recovering
process in comparison to MDP codes of the same parameters.  Even
though we are able to move in any direction with our decoding, there
exist situations where the decoder still gets lost in the middle of a
sequence because of too many erasures.  In order to restart the
decoding process one has to have access to a sufficiently large guard
space of $\nu n$ symbols.

In this section we provide a criterion (Theorem~\ref{complete-thm})
which will guarantee the computation of a guard space of sufficient
length. The special class of MDP convolutional codes which will
satisfy this assumption will be called {\bf complete MDP}
convolutional codes.

Complete-MDP convolutional
codes will turn out to be both MDP convolutional codes and reverse 
MDP codes. If the decoder gets lost in the decoding process because
of an accumulation of too many erasures a complete MDP convolutional code 
will be able to re-start the decoding process as soon as a sequence 
of symbols is found   

These codes assume stronger conditions on the parity-check matrix of
the code which reduce the number of correct symbols per window that
one needs to observe to go back to the recovering process.  The
recovering rate per window, $R_{\omega}=\frac{\# \text{erasures
    recovered}}{\# \text{symbols in a window}}$, decreases at the
instant when it is required to compute a guard space.  the recovery
rate will be computed in for this situation in
Theorem~\ref{complete-thm}.  After a guard space is obtained, the
recovery rate is again $R_{\omega}$. The waiting time in order to
continue with the recovering process becomes shorter and we avoid the
loss of big amounts of information.

{}From now on, we make the simplified assumption that $(n-k)$ divides
the degree $\delta$ of the code, and that the code $\mathcal{C}$ has a
parity-check matrix $H(z)=H_0+H_1 z+\cdots+H_{\nu}z^{\nu}$.
Therefore, $H_{\nu}$ has full rank and $\delta=\nu(n-k)$, leading to
$L=\left\lfloor \frac{\delta}{k}\right\rfloor+\nu$.

The following matrix \small
\begin{equation}               \label{complete}
\begin{sbmatrix}{cccccc}
  H_{\nu} &  \cdots  & H_{0}  \\
          & H_{\nu } &        & H_{0}  \\
          &          & \ddots &         & \ddots \\
          &          &        & H_{\nu} & \cdots & H_0 \\
\end{sbmatrix}
\end{equation}
\normalsize will play an important role in the following. For this
reason, we will call it the \textit{partial parity-check matrix} of
the code.  Then we have the following definition.
\begin{definition}\label{complete-MDP}
  A rate $\frac{k}{n}$ convolutional code $\mathcal{C}$ with
  parity-check matrix $H(z)$ as above is called a complete-MDP
  convolutional code if in the $(L+1)(n-k)\times (\nu +L+1)n$
  partial parity-check matrix every full size minor which is
  not trivially zero, is nonzero.
\end{definition}
\begin{remark}
A full size minor formed from the columns $j_{1},j_{2},\ldots,j_{(L+1)(n-k)}$ 
is not trivially zero if and only if none of these conditions is violated
\begin{itemize}
   \item $j_{s(n-k)+1}>sn$
   \item $j_{s(n-k)}\leq sn  +\nu n$
\end{itemize}
for $s=1,2,\ldots,L$.
\end{remark}
Based on many small examples,
like Example~\ref{examplecompleteMDP}, we conjecture the
existence of complete-MDP convolutional codes for every set of
parameters.  
\begin{example}\label{examplecompleteMDP}
Let  $H(z)$ be a parity-check matrix of  a $(3,1,1)$ convolutional code over $\mathbb{F}_{128}$ 
  \[
  H(z)=\begin{sbmatrix}{ccc}
    \alpha^{76}+\alpha^{77}z & \alpha^{62}+\alpha^{85}z & 1+\alpha^{76}z \\
    \alpha^{73}+\alpha^{37}z & \alpha^{76}+\alpha^{77}z &
    \alpha^{62}+\alpha^{85}z
  \end{sbmatrix}, 
  \] where 
  $\alpha^{7}+\alpha^{6}+\alpha^{3}+\alpha+1=0$.
  Note that in this case $n-k=2$ does not  divide $\delta=1$.
  The partial parity-check matrix satisfies  the condition that all
  its full size minors that are non trivially zero, that is, the
  ones that do not include columns $1$, $2$ and $3$ or $7$, $8$
  and $9$, are nonzero.
  \[
  \begin{sbmatrix}{ccccccccc}
    \alpha^{77} & \alpha^{85} & \alpha^{76} & \alpha^{76} & \alpha^{62} & 1 &0&0&0\\
    \alpha^{13} & \alpha^{77} & \alpha^{85} & \alpha^{73} & \alpha^{76} & \alpha^{82}&0&0&0 \\
   0 &  0           & 0          & \alpha^{77} & \alpha^{85} &
    \alpha^{76} & \alpha^{76} & \alpha^{62} & 1 \\
    0&     0        &  0         & \alpha^{13} & \alpha^{77} &
    \alpha^{85} & \alpha^{73} & \alpha^{76} & \alpha^{82} \\
  \end{sbmatrix}.
  \]Therefore this code is  complete-MDP.
  \hfill$\square$
 
\end{example}

\medskip

\begin{lemma}
  Every complete-MDP convolutional code is reverse-MDP. In  
  particular, every complete-MDP is an MDP code. 
\end{lemma}
\begin{IEEEproof} 
The claim follows from the fact that the matrices 
\[
\mathcal{H}_{L}=\begin{sbmatrix}{cccc}
H_{0} \\
H_{1}  & H_{0} \\
\vdots &      & \ddots \\
H_{L}  & H_{L-1} & \cdots & H_{0}
\end{sbmatrix}, 
\]
\[
\overline{\mathcal{H}}_{L}=\begin{sbmatrix}{cccc}
H_{\nu} & H_{\nu-1} & \cdots & H_{\nu -L} \\
        & H_{\nu}   &        & H_{\nu-L+1} \\
        &           & \ddots & \vdots \\
        &           &        & H_{\nu}
\end{sbmatrix},
\]
are included in the partial parity-check matrix of the code. The
full size minors of $\mathcal{H}_{L}$ and
$\overline{\mathcal{H}}_{L}$ that are not trivially zero are also not trivially zero full size
minors of the partial parity-check matrix and hence,  by Definition
\ref{complete-MDP},  they are nonzero. Therefore, the code
is reverse-MDP.  
\end{IEEEproof}
Note that the opposite is not true in general, as we can see in the following example.
\begin{example}
  Let $H(z)$  be the parity-check matrix of a
  $(3,1,1)$ reverse-MDP convolutional code over
  $\mathbb{F}_{128}$,  \[
  H(z)=\begin{sbmatrix}{ccc}
    \alpha^{93} + \alpha^{49} z & \alpha^{19} + \alpha^{30} z 
   & \alpha^{75} + \alpha^{35} z \\
    \alpha^{61} + \alpha^{19} z & \alpha^{93} + \alpha^{49} z 
   & \alpha^{19} + \alpha^{30} z \\
  \end{sbmatrix}, \]
  where
  $\alpha^7+\alpha^6+\alpha^5+\alpha^4+\alpha^2+\alpha+1=0$.
  The code does not satisfy the complete-MDP condition because
  the columns  $1$, $5$, $6$ and $7$ of the partial parity
  check matrix
  \[
  \begin{sbmatrix}{ccccccccc}
    \alpha^{49} & \alpha^{30} & \alpha^{35} &\alpha^{93} & \alpha^{19} & \alpha^{75} &0&0&0\\
    \alpha^{19} & \alpha^{49} & \alpha^{30} &\alpha^{61} & \alpha^{93} & \alpha^{19}  &0&0&0\\
    0 & 0 & 0 & \alpha^{49} & \alpha^{30} &
    \alpha^{35} & \alpha^{93} & \alpha^{19} & \alpha^{75} \\
    0 & 0 & 0 & \alpha^{19} & \alpha^{49}
    & \alpha^{30} & \alpha^{61} & \alpha^{93} & \alpha^{19}  \\
  \end{sbmatrix}
  \]
  form a  zero minor which is not trivially zero.  \hfill$\square$
\end{example}

The use of this class of codes over the erasure channel gives some
significant improvement in the recovering process.  When we receive a
pattern of erasures that we are not able to recover, by using
complete-MDP codes, we do not need to wait until a large enough
sequence of correct symbols (a new guard space) is received.  It
suffices to have a window with a certain percentage of correct symbols
to continue the decoding process. The specific requirements on the
error pattern which allows one to compute a new guard space is given
in the following theorem:

\begin{theorem}                               \label{complete-thm}
  Given a code sequence from some complete MDP convolutional code.
  If in a window of size $(\nu+L+1)n$ there are not more than
  $(L+1)(n-k)$ erasures, and if they are distributed in such a way
  that between position $1$ and $sn$ and between positions
  $(\nu+L+1)n$ and $(\nu+L+1)n-s(n-k)$, for $s=1,2,\ldots,L+1$, there
  are not more than $s(n-k)$ erasures, then full correction of all
  symbols in this interval will be possible. In particular a new guard
  space can be computed.
\end{theorem}
\begin{IEEEproof}
  Consider the matrix introduced in\eqr{complete}. By assumption on
  the existing erasures and by the assumption that every minor
  in\eqr{complete} which is not trivially zero is nonzero, it follows
  that that all erased symbols can be uniquely computed by solving
  linear systems of equations over the base field $\F$.
\end{IEEEproof}
 
Complete-MDP convolutional codes have maximum recovering rate per
window at any instant of the process, forward and backward, since
these are both MDP and reverse-MDP codes.  When we find a pattern of
erasures that we cannot recover by forward or backward decoding, then
a guard space should be computed. The complete-MDP property guarantees
that this can be done under the relatively mild conditions of
Theorem~\ref{complete-thm}. The recovering rate per window at that
instant decreases from $R_{\omega}=\frac{(L+1)(n-k)}{(L+1)n}$ to
$R_{\omega}=\frac{(L+1)(n-k)}{(L+1+\nu)n}$, since we need to observe a
bigger amount of correct information.

The following example points out the kind of situations that make
these codes more powerful than MDS block codes.
\begin{example}
  Suppose that we use a $[75,50]$ MDS block code to transmit a
  sequence over an erasure channel. This code has
  $R_{\omega}=\frac{25}{75}$. Assume that we are not able to
  recover the previous blocks of the sequence,  and let the
  following be the pattern received immediately after
  \[
  \ldots\star \star\mid \overbrace{\star \star \ldots
    \star}^{(A)14} \overbrace{\vect{v}\vect{v}\ldots
    \vect{v}}^{(B)21} \overbrace{\star \star \ldots\star}^{(C)12}
  \overbrace{\vect{v}\vect{v}\ldots \vect{v}}^{(D)28}\mid
  \]
  \[
  \mid \overbrace{\vect{v}\vect{v}\ldots \vect{v}}^{(E)19}
  \overbrace{\star\star \ldots \star}^{(F)13}
  \overbrace{\vect{v}\vect{v}\ldots \vect{v}}^{(G)30}
  \overbrace{\star\star \ldots \star}^{(H)13}\mid
  \]
  \[
  \mid \overbrace{\vect{v}\vect{v}\ldots \vect{v}}^{(I)30}
  \overbrace{\star\star \ldots \star}^{(J)6}
  \overbrace{\vect{v}\vect{v}\ldots \vect{v}}^{(K)17}
  \overbrace{\star\star\ldots \star}^{(L)22}\mid \star \star
  \ldots
  \]
  In this case the block code can not recover any of these erasures,
  thus missing $80$ information symbols.

  Note that if we use an MDP or a reverse-MDP convolutional code
  with parameters $(3,2,16)$, %in order to have the same
 % recovering capability per window, 
  we would not  be able to
  recover these erasures either, since one cannot find enough guard space, 
  of  at least $48$ correct symbols, in between the bursts. 

  Assume now that we use a $(3,2,16)$ complete-MDP convolutional
  code.  The maximum recovering rate per window of this code is
  $R_{\omega}=\frac{25}{75}$, and for smaller
  window sizes is  $\frac{(j+1)(n-k)}{(j+1)n}$, $j=0,1,\ldots,23$. 
  Due to the complete-MDP property, when  lost in the decoding process,
  we can start recovering again once we find
  a window of size $(L+1+\nu)n=(24+1+16)3=123$ where not more
  than $25$ erasures occur.

   For the above pattern, a possible such window is the following
  \[
  \overbrace{\vect{v}\vect{v}\ldots \vect{v}}^{(B)21}
  \overbrace{\star \star \ldots\star}^{(C)12}
  \overbrace{\vect{v}\vect{v}\ldots \vect{v}}^{(D)28} |
  \overbrace{\vect{v}\vect{v}\ldots \vect{v}}^{(E)19}
  \overbrace{\star\star \ldots \star}^{(F)13}
  \overbrace{\vect{v}\vect{v}\ldots \vect{v}}^{(G)30}.
  \]
   Using Theorem~\ref{complete-thm} one sets up a linear system
  of equations which will recover all erasures in this interval.
  Once we have recovered this part we can go on with the next one
  \[
  \vect{v}\vect{v}\ldots \vect{v} \overbrace{\star\star \ldots
    \star}^{(H)13}\mid \overbrace{\vect{v}\vect{v}\ldots
    \vect{v}}^{(I)30}
  \]
  and finally recover
  \[
  \vect{v}\vect{v}\ldots \vect{v} \overbrace{\star\star \ldots
    \star}^{(J)6}\mid \overbrace{\vect{v}\vect{v}\ldots
    \vect{v}}^{(K)17}.
  \]
  Although we cannot recover block $A$ with $14$ erasures and
  block $L$ with $22$ we were able to recover more than $50\%$
  of the erasures in that part of the sequence, which 
  is better than what an MDS block code could  recover.  \hfill$\square$
\end{example}

\subsection{Simulations}\label{Simulations}

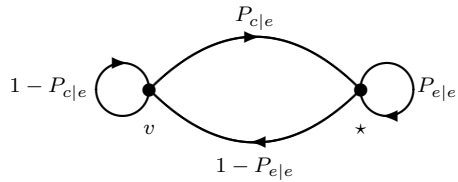
\begin{figure} 
\begin{center} 
\begin{picture}(18,8)
\thicklines
%\put(0,0){\framebox(18,8){}}

\put(5,4){\circle*{0.5}}
\put(13,4){\circle*{0.5}}

\put(4,4){\circle{2}}
\qbezier(5,4)(9,8)(13,4)

\put(14,4){\circle{2}}
\qbezier(5,4)(9,0)(13,4)

\put(9.2,6){\vector(1,0){0}}
\put(8.8,2){\vector(-1,0){0}}

\put(4.2,5){\vector(1,0){0}}
\put(13.8,3){\vector(-1,0){0}}

\put(5,2.5){\makebox(0,0){\footnotesize $v$}}
\put(13,2.5){\makebox(0,0){\footnotesize $\star$}}

\put(1.2,4){\makebox(0,0){\footnotesize $1-P_{c | e}$}}
\put(15.9,4){\makebox(0,0){\footnotesize $P_{e | e}$}}

\put(9,6.7){\makebox(0,0){\footnotesize $P_{c | e}$}}
\put(9,1){\makebox(0,0){\footnotesize $1-P_{e | e}$}}

\end{picture}
\end{center}  
\caption{Representation of the erasure channel as a Markov chain.}
\label{ECMarkov} 
\end{figure}

In this subsection we show some simulation results. Because of its
practical importance we will work with a Gilbert-Elliot channel model.
(See e.g.~\cite{mu89}). In this model the erasure probability of a
symbol is not constant and it increases after one erasure has already
occurred, in other words, the chance that another erasure occurs right
after one symbol is erased increases. We denote by $P_{c | e}$ the
probability that an erasure occurs after a correctly received symbol,
and by $P_{e | e}$ the probability that an erasure occurs after
another erasure has already happened. One way of modeling this
situation is by means of a first order Markov chain (Gilbert-Elliot
model) as shown in Figure \ref{ECMarkov}, where $0<P_{c | e}<P_{e |
  e}<1$, $\star$ represents an erasure and $v$ represents a received
symbol. In fact, Markov models are commonly used to model losses over
the Internet~\cite{ro06a}.

For these experiments we worked over erasure channels of the described
type. As we mentioned in Section \ref{Preliminaries}, the probability
that an erasure occurs after a first erasure has occurred increases,
therefore we use the following table in the simulations.
\[
\begin{tabular}{|c|c|c|c|c|c|}\hline
  $P_{c | e}$  &  $0.16$    &   $0.22$  &  $0.34$  &  $0.4$ \\  \hline
  $P_{e | e}$  &  $0.29$    &   $0.4$   &  $0.48$  &  $0.49$ \\ \hline
\end{tabular}
\]

The parameters of the codes used in the simulations are listed in the
table below, where $[N,K]$ are the parameters of an MDS block code and
$(n,k,\delta)$ the parameters used for reverse-MDP and complete-MDP
convolutional codes.

\[
\begin{tabular}{|c|c|c|c|c|c|}\hline
  Rate           &  $N$       &   $K$    &  $n$  &  $k$    & $\delta$ \\ \hline
  $2/5$  &  $100$     &   $40$   &  $5$  &  $2$    & $24$  \\ \hline
  $1/2$  &  $100$     &   $50$   &  $2$  &  $1$    & $25$  \\ \hline
  $3/5$  &  $100$     &   $60$   &  $5$  &  $3$    & $24$  \\ \hline
  $2/3$  &  $75$      &   $50$   &  $3$  &  $2$    & $16$  \\ \hline
  $7/10$ &  $100$     &   $70$   &  $10$ &  $7$    & $21$  \\ \hline
\end{tabular}
\]

\begin{figure}
  \begin{center}
    \scalebox{0.37}{\includegraphics{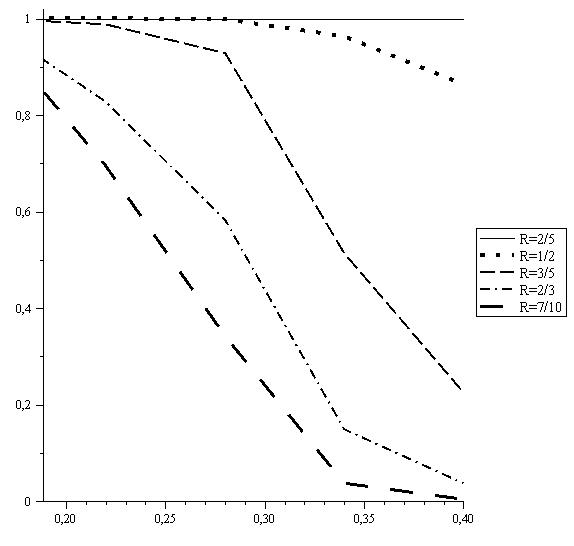} }
  \end{center}
  \caption{\small Recovering capability ($\Phi$) of MDS block
    codes with different rates in terms of the erasure
    probability of the channel ($P_{c | e}$).}
  \label{MDS_plot}
\end{figure}

Figure \ref{MDS_plot} reflects the behavior of MDS codes over the
erasure channel when choosing codes with different rates and over
channels with different erasure probabilities. The recovering
capability is expressed in terms of $\Phi=\frac{\# \text{erasures
    recovered}}{\# \text{erasures occurred}}$.

\begin{figure}
  \begin{center}
    \scalebox{0.37}{\includegraphics{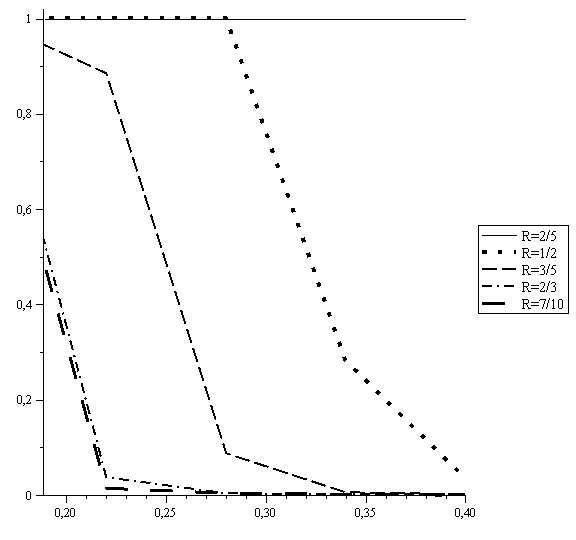} }
  \end{center}
  \caption{\small Recovering capability of reverse-MDP
    convolutional codes with different rates in terms of the
    erasure probability of the channel ($P_{c | e}$).}
  \label{reverse_plot}
\end{figure}

\begin{figure}
  \begin{center}
    \scalebox{0.37}{\includegraphics{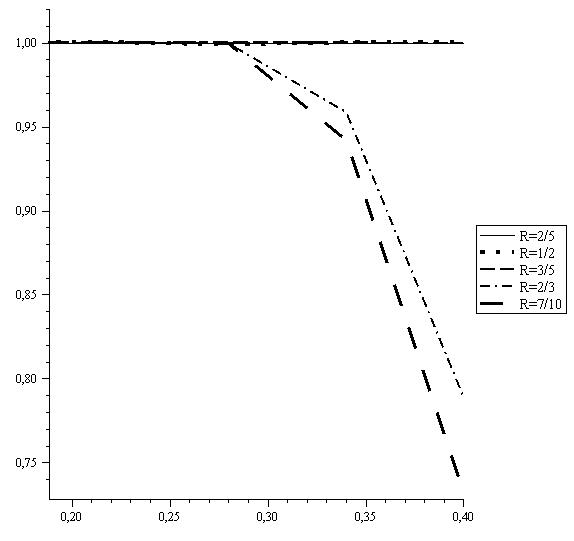} }
  \end{center}
  \caption{\small Recovering capability of complete-MDP
    convolutional codes with different rates in terms of the
    erasure probability of the channel ($P_{c | e}$).}
  \label{strong_plot}
\end{figure}

In Figures \ref{reverse_plot} and \ref{strong_plot} we can see the
performance of reveres-MDP and complete-MDP convolutional codes,
respectively. The codes were chosen to have equal transmission rate
and recovering rate per window to those of the MDS block codes used in
the simulations of Figure~\ref{MDS_plot}.
 
\normalsize

\normalsize

\normalsize

The new simulation for reverse-MDP convolutional codes shows that
reverse-MDP codes only outperform MDS codes at low rates. If we
compare Figures~\ref{MDS_plot} and \ref{reverse_plot}, one can see
that only for rates equal to $R=2/5$ and $R=1/2$ the results are
better using reverse-MDP convolutional codes.

However, observing the results in Figure~\ref{strong_plot}, one can
see how complete-MDP convolutional codes give much better performance
than MDS block codes. Even though the rate decreases for convolutional
codes when we increase the erasure probability, the behavior is better
than in the MDS case.

For this reason we propose this kind of codes as a very good
alternative to MDS block codes over this channel. Moreover, we believe
that our proposed way of generating reverse-superregular matrices in
Theorem \ref{const}, together with the construction for reverse-MDP
convolutional codes given in Section~\ref{ConstRev}, generates
complete-MDP convolutional codes; so far we did not find any evidence
of the opposite.  Unfortunately, we were not able yet to prove this
result; it remains an open question.

%%%%%%%%%%%%%%%%%%%%%%%%%%%%%%%%%%%%%%%%%%%%%%%%%%%%%%%%%%
\section{Comparison between MDS block codes and MDP convolutional codes}\label{Comparison}

As we have already pointed out through several examples MDP
convolutional codes often are capable of decoding more erasures than
comparable MDS block codes. In this section we would like to give some
theoretical results on the decoding capabilities of (complete) MDP
convolutional codes and compare these codes with MDS block codes of
the same rate.

As a first goal we will show that a rate $k/n$ convolutional
code will not be able to decode erasures at a rate of more than 
$(n-k)/n$. The following theorem serves this purpose.
\begin{theorem}                    \label{unique-dec}
Let $H(z)=\sum^{\nu}_{i=0} H_{i}z^{i}$ be the parity check matrix
of an $(n,k,\delta)$ convolutional code. Assume 
$\vec{v}(z)=\vect{v}_{0}+\vect{v}_{1}z+\ldots+\vect{v}_{l}z^{l}$
is a transmitted codeword and more than $(l+\nu+1)(n-k)$ erasures
happen during transmission. Then unique decoding is not possible.
\end{theorem} 
\begin{IEEEproof}
$\vec{v}(z)$ has to satisfy the linear system of equations as given in 
Equation\eqr{eq0}. The maximum number of erasures which uniquely can be 
decoded is hence given by the rank of the matrix appearing in Equation\eqr{eq0}.
This rank is at most  $(l+\nu+1)(n-k)$.
\end{IEEEproof}

\begin{corollary}                    \label{rec-rate}
The maximum recovery rate of an $(n,k,\delta)$ convolutional code is 
at most $\frac{n-k}{n}$.
\end{corollary}
\begin{IEEEproof}
  Theorem~\ref{unique-dec} shows that in a window of length $(l+1)n$
  at most $(l+\nu+1)(n-k)$ erasures can be decoded. Taking the limit
  $l\longrightarrow\infty$ we see that not more than a ratio of
  $\frac{n-k}{n}$ erasures can be decoded.
\end{IEEEproof}

As a result we see that for long messages a rate 
$k/n$ convolutional code cannot decode at a rate larger than
$(n-k)/n$. On the other hand we have seen in Corollary~\ref{main} 
that an MDP convolutional code can decode all erasures 
as long as there are at most $(L+1)(n-k)$ in any sliding
window of length $(L+1)n$. 

Compare this now with an $[N,K]$ linear block code $\C$. 
The maximum number of erasures which can be decoded in
any block of length $N$ is $N-K$ and this maximum is achieved 
by an MDS block code of rate $K/N$. As a consequence the
recovery rate of a rate $k/n$ MDP convolutional code and a
rate $k/n$ block code are therefore the same `on average'. 
What matters for block codes is the block length and 
what matters for convolutional codes is the degree. 

We conclude the section by comparing a $(2,1,\delta)$ 
convolutional code with an $[N,K]=[2\delta,\delta]$ block code.

Both these codes have rate $1/2$. The $[2\delta,\delta]$ block code
can decode all erasures as long as there are at most 
$\delta$ erasures in every slotted window (=block) of length $2\delta$.

The performance of a $(2,1,\delta)$ (complete) MDP convolutional code 
is as follows:

By Corollary~\ref{main} unique decoding from left to right 
is possible as long as there are at most $(\delta +1)$ erasures in 
any sliding window of length $2\delta +2$. If the  $(2,1,\delta)$
code is also a complete MDP convolutional code then 
Theorem~\ref{complete-thm} states that decoding a whole window of
 length $6\delta +2$ can be achieved as long as there are not
more than $2\delta +1$ erasures, and these erasures do not
concentrate on the boundaries of the interval.
In this way guard spaces can be computed and full decoding 
is possible via the forward, backward decoding process as we 
described it at length before.

The comparison shows that in order that a block
code of rate $1/2$ can compete with a $(2,1,\delta)$
MDP convolutional code a block length of at least 
$2\delta$ is needed and even then there are 
many situations where full decoding is possible
with the convolutional code and blocks of the 
linear block code cannot be decoded.

We conclude the section by comparing the decoding complexity.

An $[N,K]$ MDS block is capable of decoding $N-K$ erasures
in every block. Assume $N-K$ erasures actually happen. 
If one works with the parity check matrix then the decoding
task naturally translates into a linear system of the form
$Ax=b$, where $A$ is an $(N-K)\times (N-K)$ consisting
of the columns of the parity check matrix where the erasures
actually did happen. Alternatively one can work with the
generator matrix of the code and again ends up with a linear
system of the form $Ax=b$, where $A$ is a $K\times K$ matrix
consisting of the $K$ columns of the generator matrix where
the transmission arrived correctly. 

The number of field operations required to decode is hence of the 
order $O(r^3)$, where $r=\min\{ K,N-K\}$. 

For an $(n,k,\delta)$ convolutional code the iterative decoding
process as described in Theorem~\ref{main2} requires again the
solution of a linear system of the form $Ax=b$, where $A$ is in the
worst case of size $(L+1)(n-k)\times (L+1)(n-k)$, in case one works
with the parity check matrix $H(z)$. If the number of erasures is
relatively mild (always less than $(L+1)(n-k)$ erasures in any sliding
window of length $(L+1)n$) then each system of equations of the form
$Ax=b$ will decode one to several erasures at the time. If more
erasures accumulate then Theorem~\ref{complete-thm} has to be invoked
which requires the solution of a linear system $Ax=b$ of slightly
larger size and this system possibly recovers just one erasure.

If $n(L+1)$ is comparable to the block length $N$ of the MDS block
code then one sees that the computational effort is very comparable.

%%%%%%%%%%%%%%%%%%%%%%%%%%%%%%%%%%%%%%%%%%%%%%%%%%%%%%%%%%
\section{Conclusions}\label{Conc}

In this paper, we propose MDP convolutional codes as an alternative to
MDS block codes when decoding over an erasure channel.  MDP
convolutional codes can be decoded iteratively `from left to right' as
long as the number of erasures in any sliding window does not surpass
a certain amount (Corollary~\ref{main}).

Reverse MDP convolutional codes are MDP convolutional codes having
the extra property that erasures can also be decoded `from right to
left' as long as the number of erasures in any sliding window does not
surpass a certain amount (Theorem~\ref{main-rev}).

Complete MDP convolutional codes are reverse MDP convolutional codes
having the additional property that a whole interval can be decoded
(independent of the past and the future) as long as the number of
erasures does not surpass a certain amount
(Theorem~\ref{complete-thm}).

The maximum erasure recovery rate of a rate $k/n$  MDP convolutional code
is $\frac{n-k}{n}$. This is the same recovery rate as for a rate $k/n$ 
MDS block code often used in practice. In the case of an $[N,K]$ MDS block code
error free decoding is possible if in every block at most $N-K$ erasures 
do happen. An $(n,k,\delta)$ MDP convolutional code can perform
error free communication if in every sliding window of length 
$n(L+1)$ at most $(n-k)(L+1)$ errors do happen, where 
$ L=\left\lfloor \frac{\delta}{k} \right\rfloor + 
  \left\lfloor \frac{\delta}{n-k} \right\rfloor$.
When $N=nL$ then an MDS $[N,K]$ block code is comparable to an
MDP convolutional code of the same rate. However simulation results 
show that even in this situation MDP convolutional codes perform better 
in case the convolutional code is a complete MDP convolutional code.

\section*{Acknowledgment}

We would like to thank Martin Haenggi for explaining us the
distribution of packet sizes when transmitting files over the
Internet. The authors are also grateful to the anonymous referees for
the many insightful comments they provided.

% \bibliography{huge}\bibliographystyle{plain}\end{document}

\def\cprime{$'$} \def\polhk#1{\setbox0=\hbox{#1}{\ooalign{\hidewidth
      \lower1.5ex\hbox{`}\hidewidth\crcr\unhbox0}}}
\def\polhk#1{\setbox0=\hbox{#1}{\ooalign{\hidewidth
      \lower1.5ex\hbox{`}\hidewidth\crcr\unhbox0}}} \def\cprime{$'$}
\def\cprime{$'$} \def\cprime{$'$} \def\cprime{$'$}

\begin{IEEEbiographynophoto}{Virtudes Tom\'{a}s}
  was born in Spain in 1983.  She received her B.A. in Mathematics in
  2006 from the University of Alicante with an Extraordinary Award.
  {}In 2010 she obtained the Ph.D. degree from the University of
  Alicante and her dissertation was supervised by Prof.  Joan-Josep
  Climent and Prof.  Joachim Rosenthal. Her thesis is in Coding Theory
  and its main topic is concerned with Complete-MDP convolutional
  codes.

  During her Ph.D. studies she was supported by an FPU Grant from the
  regional government of La Generalitat Valenciana (research grant for
  Ph.D. students) and enjoyed two research visits abroad, one in 2008
  when she spent 12 months as a visitor at the University of Z\"urich
  (Z\"urich, Switzerland) and a second one in 2009 when she visited San
  Diego State University (San Diego, USA) for 2 months.
\end{IEEEbiographynophoto}

\begin{IEEEbiographynophoto}{Joachim Rosenthal}
  received the Diplom in Mathematics from the University of Basel in
  1986 and the Ph.D. in Mathematics from Arizona State University in
  1990.  Since 2004 he has been Professor of Applied Mathematics at
  the University of Z\"urich where he currently also serves as
  Director of the Mathematics Institute.

  From 1990 until 2006 he has been with the University of Notre Dame,
  where he has last been the holder of an endowed chair in Applied
  Mathematics and also was Concurrent Professor in Electrical
  Engineering.

  In the academic year 1994/1995 he spent a sabbatical year at CWI the
  Center for Mathematics and Computer Science in Amsterdam, The
  Netherlands. During the academic year 1999/2000 he was a Guest
  Professor at the Swiss Federal Institute of Technology in Lausanne,
  Switzerland, affiliated with the School of Computer \& Communication
  Sciences.

  His current research interests are in coding theory and
  cryptography.  He currently serves as Associate Editor for {\em
    Journal of Algebra and its Applications} (JAA) and {\em Advances
    in Mathematics of Communications} (AMC). In the past he served
  also on the editorial boards of {\em SIAM Journal on Control and
    Optimization} (SICON), {\em Mathematics of Control, Signals, and
    Systems} (MCSS), {\em Linear Algebra and its Applications} (LAA)
  and {\em Journal of Mathematical Systems, Estimation, and Control}.
  In 2002 he served as the symposium chair of the International
  Symposium on Mathematical Theory of Networks and Systems (MTNS) and
  in 2010 he served together with M.  Greferath as conference chair of
  the IEEE Information Theory Workshop in Dublin.
\end{IEEEbiographynophoto}

\begin{IEEEbiographynophoto}{Roxana Smarandache}
  is an associate professor in the Department of Mathematics and
  Statistics at San Diego State University. Originally from Bucharest,
  Romania, she has completed her undergraduate studies in mathematics
  at the University of Bucharest in 1996, with a B.S.  thesis on
  Number Theory.  From 1996-2001 she pursued a Ph.D. degree in
  Mathematics at the University of Notre Dame, which she completed in
  July 2001.  Her thesis is in Coding Theory, with the subject of
  algebraic convolutional codes. After her Ph.D. she joined San Diego
  State University.

  During the academic year 1999-2000, Dr. Smarandache was for six
  months a visiting scholar at the Swiss Federal Institute of
  Technology (EPFL), Switzerland, in the Department of Communication
  Systems. During the academic year 2005-2006, she was on leave at the
  University of Notre Dame, on a visiting assistant professor position
  in the Department of Mathematics.  During the academic year
  2008-2009, she spent part of a sabbatical year at the University of
  Zurich (8 months) and part at the University of Notre Dame (3
  months).

  Dr. Smarandache's research topics are mainly related to coding
  theory.  Her recent interests include low density parity check
  codes, iterative and linear programming decoding, and convolutional
  codes.
\end{IEEEbiographynophoto}

\end{document}